\documentclass[twocolumn,showpacs,prc,reprint,nofootinbib,a4paper]{revtex4}  
\usepackage{epsfig}
\usepackage[english]{babel}

\newcommand{\ee}{$e^{+}e^{-}$}

\newcommand{\pnb}{p+$^{93}$Nb}

\newcommand{\gevu}{GeV/$u$}

\newcommand{\gevcc}{GeV/$c^{2}$}

\newcommand{\gevc}{GeV/$c$}

\newcommand{\tev}{TeV}
\newcommand{\gev}{GeV}
\newcommand{\mev}{MeV}
\newcommand{\etal}{$et~al.$}
\newcommand{\eeee}{e$^+$e$^-$e$^+$e$^-$}

\begin{document}

\title{Inclusive pion and eta production in p+Nb collisions at 3.5 \gev\ beam energy}


\author{G.~Agakishiev$^{7}$, A.~Balanda$^{3}$, D.~Belver$^{18}$, A.~Belyaev$^{7}$, 
J.C.~Berger-Chen$^{9}$, A.~Blanco$^{2}$, M.~B\"{o}hmer$^{10}$, J.~L.~Boyard$^{16}$, P.~Cabanelas$^{18}$, 
S.~Chernenko$^{7}$, A.~Dybczak$^{3}$, E.~Epple$^{9}$, L.~Fabbietti$^{9}$, O.~Fateev$^{7}$, 
P.~Finocchiaro$^{1}$, P.~Fonte$^{2,b}$, J.~Friese$^{10}$, I.~Fr\"{o}hlich$^{8}$, T.~Galatyuk$^{5,c}$, 
J.~A.~Garz\'{o}n$^{18}$, R.~Gernh\"{a}user$^{10}$, K.~G\"{o}bel$^{8}$, M.~Golubeva$^{13}$, D.~Gonz\'{a}lez-D\'{\i}az$^{5}$, 
F.~Guber$^{13}$, M.~Gumberidze$^{5,16,*}$, T.~Heinz$^{4}$, T.~Hennino$^{16}$, R.~Holzmann$^{4,*}$, 
A.~Ierusalimov$^{7}$, I.~Iori$^{12,e}$, A.~Ivashkin$^{13}$, M.~Jurkovic$^{10}$, B.~K\"{a}mpfer$^{6,d}$, 
T.~Karavicheva$^{13}$, I.~Koenig$^{4}$, W.~Koenig$^{4}$, B.~W.~Kolb$^{4}$, G.~Kornakov$^{18}$, 
R.~Kotte$^{6}$, A.~Kr\'{a}sa$^{17}$, F.~Krizek$^{17}$, R.~Kr\"{u}cken$^{10}$, H.~Kuc$^{3,16}$, 
W.~K\"{u}hn$^{11}$, A.~Kugler$^{17}$, A.~Kurepin$^{13}$, V.~Ladygin$^{7}$, R.~Lalik$^{9}$, 
S.~Lang$^{4}$, K.~Lapidus$^{9}$, A.~Lebedev$^{14}$, T.~Liu$^{16}$, L.~Lopes$^{2}$, 
M.~Lorenz$^{8,c}$, L.~Maier$^{10}$, A.~Mangiarotti$^{2}$, J.~Markert$^{8}$, V.~Metag$^{11}$, 
B.~Michalska$^{3}$, J.~Michel$^{8}$, C.~M\"{u}ntz$^{8}$, L.~Naumann$^{6}$, Y.~C.~Pachmayer$^{8}$, 
M.~Palka$^{3}$, Y.~Parpottas$^{15,f}$, V.~Pechenov$^{4}$, O.~Pechenova$^{8}$, J.~Pietraszko$^{4}$, 
W.~Przygoda$^{3}$, B.~Ramstein$^{16}$, A.~Reshetin$^{13}$, A.~Rustamov$^{8}$, A.~Sadovsky$^{13}$, 
P.~Salabura$^{3}$, A.~Schmah$^{9,a}$, E.~Schwab$^{4}$, J.~Siebenson$^{9}$, Yu.G.~Sobolev$^{17}$, 
S.~Spataro$^{11,g}$, B.~Spruck$^{11}$, H.~Str\"{o}bele$^{8}$, J.~Stroth$^{8,4}$, C.~Sturm$^{4}$, 
A.~Tarantola$^{8}$, K.~Teilab$^{8}$, P.~Tlusty$^{17}$, M.~Traxler$^{4}$, R.~Trebacz$^{3}$, 
H.~Tsertos$^{15}$, T.~~Vasiliev$^{7}$, V.~Wagner$^{17}$, M.~Weber$^{10}$, C.~Wendisch$^{6,d}$, 
J.~W\"{u}stenfeld$^{6}$, S.~Yurevich$^{4}$, Y.~Zanevsky$^{7}$}

\affiliation{
(HADES collaboration) \\\mbox{$^{1}$Istituto Nazionale di Fisica Nucleare - Laboratori Nazionali del Sud, 95125~Catania, Italy}\\
\mbox{$^{2}$LIP-Laborat\'{o}rio de Instrumenta\c{c}\~{a}o e F\'{\i}sica Experimental de Part\'{\i}culas , 3004-516~Coimbra, Portugal}\\
\mbox{$^{3}$Smoluchowski Institute of Physics, Jagiellonian University of Cracow, 30-059~Krak\'{o}w, Poland}\\
\mbox{$^{4}$GSI Helmholtzzentrum f\"{u}r Schwerionenforschung GmbH, 64291~Darmstadt, Germany}\\
\mbox{$^{5}$Technische Universit\"{a}t Darmstadt, 64289~Darmstadt, Germany}\\
\mbox{$^{6}$Institut f\"{u}r Strahlenphysik, Helmholtz-Zentrum Dresden-Rossendorf, 01314~Dresden, Germany}\\
\mbox{$^{7}$Joint Institute of Nuclear Research, 141980~Dubna, Russia}\\
\mbox{$^{8}$Institut f\"{u}r Kernphysik, Goethe-Universit\"{a}t, 60438 ~Frankfurt, Germany}\\
\mbox{$^{9}$Excellence Cluster 'Origin and Structure of the Universe' , 85748~Garching, Germany}\\
\mbox{$^{10}$Physik Department E12, Technische Universit\"{a}t M\"{u}nchen, 85748~Garching, Germany}\\
\mbox{$^{11}$II. Physikalisches Institut, Justus Liebig Universit\"{a}t Giessen, 35392~Giessen, Germany}\\
\mbox{$^{12}$Istituto Nazionale di Fisica Nucleare, Sezione di Milano, 20133~Milano, Italy}\\
\mbox{$^{13}$Institute for Nuclear Research, Russian Academy of Science, 117312~Moscow, Russia}\\
\mbox{$^{14}$Institute of Theoretical and Experimental Physics, 117218~Moscow, Russia}\\
\mbox{$^{15}$Department of Physics, University of Cyprus, 1678~Nicosia, Cyprus}\\
\mbox{$^{16}$Institut de Physique Nucl\'{e}aire (UMR 8608), CNRS/IN2P3 - Universit\'{e} Paris Sud, F-91406~Orsay Cedex, France}\\
\mbox{$^{17}$Nuclear Physics Institute, Academy of Sciences of Czech Republic, 25068~Rez, Czech Republic}\\
\mbox{$^{18}$LabCAF. F. F\'{\i}sica, Univ. de Santiago de Compostela, 15706~Santiago de Compostela, Spain}\\ 
\\
\mbox{$^{a}$ now at Lawrence Berkeley National Laboratory, ~Berkeley, USA}\\
\mbox{$^{b}$ also at ISEC Coimbra, ~Coimbra, Portugal}\\
\mbox{$^{c}$ also at ExtreMe Matter Institute EMMI, 64291~Darmstadt, Germany}\\
\mbox{$^{d}$ also at Technische Universit\"{a}t Dresden, 01062~Dresden, Germany}\\
\mbox{$^{e}$ also at Dipartimento di Fisica, Universit\`{a} di Milano, 20133~Milano, Italy}\\
\mbox{$^{f}$ also at Frederick University, 1036~Nicosia, Cyprus}\\
\mbox{$^{g}$ now at Dipartimento di Fisica Generale and INFN, Universit\`{a} di Torino, 10125~Torino, Italy}\\
\mbox{$^{*}$ Corresponding authors: R.Holzmann@gsi.de, M.Gumberidze@gsi.de}\\
}

\date{\today}

\begin{abstract}
Data on inclusive pion and eta production measured with the dielectron spectrometer
HADES in the reaction \pnb\ at a kinetic beam energy of 3.5 \gev\ are presented.
Our results, obtained with the photon-conversion method, supplement the rather sparse
information on neutral-meson production in proton-nucleus reactions existing for this
bombarding energy regime.  The reconstructed \eeee\ transverse-momentum and rapidity
distributions are confronted with transport-model calculations, which account fairly well
for both $\pi^0$ and $\eta$ production.

\end{abstract}


\maketitle

\section{Introduction}
\label{intro}

The High-Acceptance DiElectron Spectrometer (HADES) experiment at GSI pursues
a comprehensive program of dielectron emission studies in few-\gev\
nucleon-nucleon \cite{hades_p35p,hades_p22p}, proton-nucleus \cite{hades_pNb},
and nucleus-nucleus collisions \cite{hades_arkcl}. 
Dilepton spectroscopy allows to investigate the properties of hadrons
produced, propagated, and decayed in a strongly interacting medium.  This is because
leptons (electrons and muons) do not themselves interact strongly when traveling
through finite-sized hadronic matter, that is, their kinematics remain basically undistorted.
Lepton-pair measurements are hence ideally suited to search for medium modifications of
hadrons in nuclear matter \cite{LeupoldMetagMosel,HayanoHatsuda}.  The observed dilepton
spectra consist, however, of a complex superposition of various mesonic and baryonic
contributions, and their interpretation requires a detailed knowledge of all sources.
Indeed, early interpretations of dilepton spectra from relativistic heavy-ion collisions
commonly introduced a schematic distinction of (i) hard initial contributions related to
Drell-Yan type processes, (ii) the thermal radiation off the fireball, and (iii) the
hadronic cocktail from late decays following its disassembly (cf.~\cite{kaempfer}).
Transport models supersede this artificial separation as they describe all phases of
the collision on an equal footing by following continuously virtual and real photon
emission over time.  Presently, they are commonly employed in the few-\gev\ bombarding
energy regime to describe particle production and propagation through the medium,
in particular, when dealing with the complex dynamics of nucleus-nucleus reactions
\cite{hsd1,hsd2,urqmd1,urqmd2,gibuu1,gibuu2}.  Comprehensive information on meson
production is thereby mandatory to benchmark and constrain those calculations.
In this context the neutral pion and eta mesons are of particular interest as they
contribute largely to the dilepton spectrum via their Dalitz decays,
$\pi^0 \rightarrow \gamma\gamma^* \rightarrow \gamma$\ee\
and  $\eta \rightarrow \gamma\gamma^* \rightarrow \gamma$\ee, respectively.

Although in the few-\gev\ energy regime a large body of systematic data on pion and,
to a lesser extent, on eta production in nucleus-nucleus collisions has been gathered over
the last decades, mostly at the Bevalac, the AGS, and SIS18 accelerators, there is much
less information available from proton-nucleus reactions.  The latter ones are, however,
important as an intermediate step between nucleon-nucleon and nucleus-nucleus collisions.
Charged pions from p+A have been measured at the Bevalac with proton beams
of kinetic energies up to 2.1~\gev\ \cite{nagamiya}, at TRIUMF up to 0.5~\gev\ \cite{digiacomo},
and recently by the HARP experiment at the CERN PS with proton energies between
2~and 12~\gev\ \cite{harp}.  Information on eta production in p+A reactions is even
more scarce.  Only the PINOT experiment at the SATURNE accelerator in Saclay provided
data for proton energies in the range of 0.8 -- 1.5~\gev\ \cite{pinot}.

In this paper we supplement the available body of experimental results on pion ($\pi^0, \pi^-$)
and eta production in p+A collisions with data obtained with HADES in the p+Nb reaction at 3.5~\gev.
Negative pions have been identified via their characteristic energy loss vs. momentum signature
in the HADES time-of-flight (TOF) system.  The neutral mesons, $\pi^0$ and $\eta$, were reconstructed
with the photon-conversion technique in which meson decay photons are detected via their external
conversion into an \ee\ pair via a Bethe-Heitler process, preferentially in high-$Z$ materials.
This method has been developed foremost in high-energy physics
experiments \cite{expE672,expE771,expCDF,expATLAS,expCMS} to study the radiative decays
of the quarkonium states $\chi_c$ into $J/\Psi + \gamma$ and $\chi_{b}$ into $\Upsilon + \gamma$.
It has also been used in high-energy heavy-ion reactions, namely by the \mbox{PHENIX} experiment
at RHIC, studying Au+Au collisions at a nucleon-nucleon center-of-mass energy of
$\sqrt{s_{NN}}=200$~\gev\ \cite{phenix} and by the \mbox{ALICE} experiment at the LHC
in $\sqrt{s_{NN}}=7$~\tev\ p+p collisions \cite{alice}.  Making use of the good momentum resolution
of charged-particle trackers, in particular at low energies, the conversion technique
offers typically better energy resolution than a photon calorimeter.
We demonstrate here the applicability of the method with HADES in few-GeV reactions.

Our paper is organized as follows.  Section~\ref{experiment} describes the experiment and the employed
particle identification procedures.  In Sec.~\ref{method} the photon-conversion method is introduced.
Pion and eta spectra, as well as meson multiplicities are presented in Sec.~\ref{yields}. 
In Sec.~\ref{transport} we compare the data with transport-model calculations and, finally,
in Sec.~\ref{summary} we summarize our findings.  A preliminary version of the $\pi^-$ data
shown here has already been presented elsewhere \cite{hades_pim}.

\section{The experiment}
\label{experiment}

The six-sector high-acceptance spectrometer HADES operates at the
GSI Helmholtzzentrum f\"{u}r Schwerionenforschung in Darmstadt where it
takes beams from the heavy-ion synchrotron SIS18.  Although its setup
was originally optimized for dielectron spectroscopy, HADES is in
fact a versatile charged-particle detector with both good efficiency and
momentum resolution.  Its main component serving for electron and positron
selection is a hadron-blind Ring-Imaging Cherenkov detector (RICH).
Further particle identification power is provided by the time of flight
measured in a plastic scintillator TOF wall, the electromagnetic shower
characteristics observed in a pre-shower detector, and the energy-loss
signals from the scintillators of the TOF wall as well as from the four
planes of drift chambers serving as tracking stations.  Charged particles
are tracked through a toroidal magnetic field provided by a six-coil
iron-less superconducting magnet.  All technical aspects of the detector
are described in~\cite{hades_tech}.

In the experiment discussed here a proton beam with a kinetic energy
of $E_p$ = 3.5~\gev\ and an average intensity of about $2\times10^{6}$
particles per second impinged onto a 12-fold segmented niobium ($^{93}$Nb) target
with a total thickness of 5.4~mm and corresponding to 2.8\% nuclear interaction
probability.  The online event selection was done in two steps:
a 1$^{st}$-level trigger (LVL1) selected events with at least three charged-particle
hits in the TOF wall ($N_{ch}\ge3$) and a 2$^{nd}$-level trigger (LVL2) fired if
an electron or positron candidate was recognized.  While all LVL2 events
were recorded, the LVL1-only events were downscaled by a factor three
before being written to data storage.  This trigger scheme was in fact
primarily optimized for studying inclusive \ee\ production \cite{hades_pNb}.
To allow for trigger bias studies also LVL1 events requiring only two charged particles
were recorded during part of the experiment.  The LVL1 triggered on p+Nb reactions
with 56\% efficiency in the $N_{ch}\ge3$ mode and 72\% efficiency in the $N_{ch}\ge2$ mode.
These efficiencies correspond to the fraction of reactions that actually fired the LVL1
trigger.  In total $4.6\times10^{9}$ events -- downscaled LVL1 or LVL2 -- were recorded,
corresponding to $7.7\times10^{9}$ inspected LVL1 events and $1.3\times10^{10}$ reactions
in the target. 

To study neutral-meson production the off-line data analysis searched for events
containing four lepton tracks from which the 4-momentum of $\pi^0$ and $\eta$ mesons
was fully reconstructed.  Indeed, the electromagnetic decays of the latter,
that is mostly $\pi^0,\eta \rightarrow \gamma$e$^+$e$^-$ (Dalitz) and
$\pi^0,\eta \rightarrow \gamma\gamma$, combined with the external conversion
of the decay photon(s) on material in the inner region of the HADES detector
lead to events with a characteristic 4-lepton signature, namely
$\pi^0, \eta \rightarrow$ \eeee.  Simulations show that the Dalitz decay
contributes 25\% (30\%) of the detectable $\pi^0$ ($\eta$) yield.
The direct branching into \eeee\ due to the decay into two virtual photons
($\pi^0, \eta \rightarrow \gamma^*\gamma^*$) is however very small \cite{pdg2012}
and contributes in our case only about 2 -- 2.5\% of the total yield.
Electron and positron tracks were identified following the procedures described
in detail in \cite{hades_tech,hades_arkcl}.  As no dedicated start
detector was present in this experimental run, the start time for the time-of-flight
measurement was reconstructed event-by-event from the most optimal assignment
of different particle hypotheses (e$^-$ or $\pi^-$) to tracks of negatively charged fast particles.
In the extraction of the inclusive negative pion yields no direct use was however
made of the time-of-flight information, and the pion identification was solely
based on energy-loss vs. momentum cuts (for details see \cite{hades_pim}).
Positive pions were identified likewise but, their spectra being partially
contaminated by the much more abundant protons, they were not analyzed further.

\section{The photon-conversion method}
\label{method}

The detection of high-energy photons via a full reconstruction of conversion pairs
has been applied with success by the PHENIX \cite{phenix} and ALICE \cite{alice} collaborations
in ultrarelativistic collisions.  Although those experiments comprise also electromagnetic
calorimeters for photon detection, this alternative method is considered to offer valuable
supplementary information on thermal photon as well as neutral-meson production.
As HADES is presently not equipped with such a calorimeter, the conversion method
opens a unique approach to $\pi^0$ and $\eta$ detection.

Meson reconstruction was realized with identified lepton tracks by joining
opposite-sign leptons into \ee\ pairs and by further combining those dileptons
pairwise into \eeee\ multiplets.  Calculating the four-lepton
invariant mass $M_{e^+e^-e^+e^-}$ and setting appropriate mass cuts allows to select
the $\pi^0$ and $\eta$ mesons, respectively.  Various opening angle cuts, optimized
in extensive Monte-Carlo simulations,  were applied
to suppress combinations of uncorrelated leptons, namely $\theta_{e^+e^-} < 2.5^{\circ}$ on
the dilepton with the smaller opening angle and $\theta_{e^+e^-} < 20^{\circ}$ on the second one
(all angles given in the laboratory system).  The $2.5^{\circ}$ selection is optimal for
conversion pairs while the $20^{\circ}$ cut accepts also the more massive and hence wider Dalitz pairs.
In addition, a cut was applied on the relative angle between the two dileptons in a multiplet,
namely $\theta_{\gamma*\gamma*}>5^{\circ}$, to suppress spurious counts at low $M_{e^+e^-e^+e^-}$.
These cuts were adjusted on the data in order to maximize the meson yield while keeping
the background of uncorrelated combinations low.

\begin{figure}[!htb]

  \mbox{\epsfig{width=0.99\linewidth, height=0.99\linewidth, figure=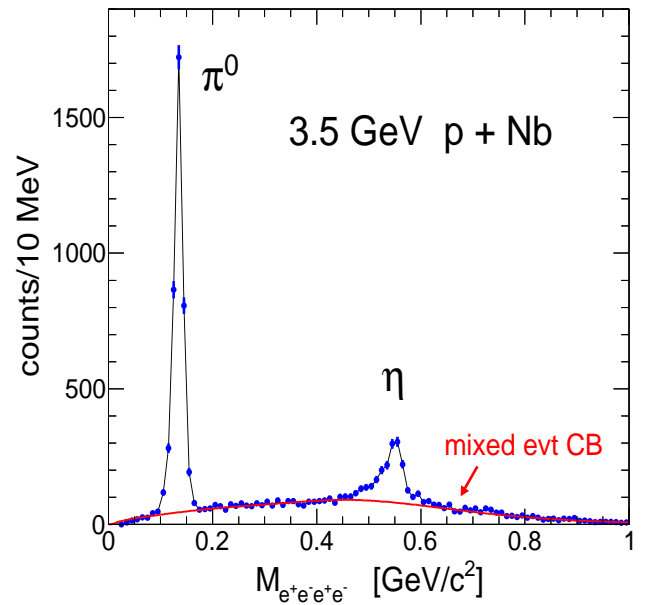}}
  \vspace*{-0.2cm}
  \caption[] {(Color online) Invariant-mass distribution of all \eeee\ multiplets
    (dots, error bars are statistical) measured with HADES in the 3.5~\gev\ p+Nb reaction.
    To improve visibility, the data points are connected by a thin curve.  The background
    of uncorrelated lepton combinations obtained from event mixing is shown as well (solid curve).
    }
  \vspace*{-0.2cm}
  \label{Meeee}
\end{figure}

\begin{table}[htb]

\caption[]{Characteristics of the reconstructed meson peaks: raw signal counts above CB (integrated
           in the range 0.10 -- 0.16~\gevcc\ for $\pi^0$ and 0.46 -- 0.60~\gevcc\ for $\eta$),
           signal/CB ratio in those mass ranges, position of the peak maximum, and $\sigma$ width
           of the peak ($\sigma =$FWHM/2.35).  All errors are statistical.
          }      

\vspace*{0.2cm}
\begin{center}
\begin{tabular}{ l  c c } 
  \hline
  \hline
  identified meson~ & ~$\pi^0$ &  $\eta$ \\
  \hline
  signal  [counts]~ & $3800\pm63$ & $1240\pm49$ \\
  signal/CB~ & ~18.1 & 1.1 \\
  position  [MeV]~~ & $134\pm1$ & $547\pm2$ \\
  width  [MeV]~~ & $8.0\pm0.6$ & $19\pm2$ \\
  \hline
  \hline

\end{tabular}
\end{center}
\label{peaktable}
\end{table}

The resulting $M_{e^+e^-e^+e^-}$ distribution is shown in Fig.~\ref{Meeee}.
Prominent peaks appear at the masses of the $\pi^0$ and the $\eta$ mesons on top of
a continuum yield attributable to combinations of uncorrelated leptons.  A systematic
investigation of this combinatorial background (CB) was done, both by comparing the mass
distributions of various charge combinations other than + -- + -- and by applying an
event mixing technique.  We find that the largest part ($>$~90\%) of the CB originates
from the combination of uncorrelated dileptons, that is uncorrelated photons
(real or virtual), mostly from multi-$\pi^0$ events.
In that sense, the \eeee\ CB behaves very much like the two-photon
CB observed in a calorimeter and it can hence be determined by event mixing.
Thus, after subtraction of a mixed event CB normalized in the peak-free regions of the
mass spectrum, the $\pi^0$ and $\eta$ peak characteristics are straightforwardly extracted.
Integrated raw yields, peak positions and widths (defined as $\sigma=$FWHM/2.35)
are listed in Table~\ref{peaktable}.  Systematic errors on the yields from CB subtraction,
estimated by varying the weights of the two normalization regions 0.2--0.4 and 0.6--1.0~\gevcc,
are 5\% for the $\pi^0$ and 10\% for the $\eta$, respectively.  Having corrected
the individual lepton momenta for their energy loss of typically 2--3~\mev, both peak positions
are found to be consistent with the nominal meson masses, namely $M_{\pi^0}=0.13498$~\gevcc\ 
and $M_{\eta}=0.54785$~\gevcc\ \cite{pdg2012}.  The peak widths are determined by
the momentum resolution of the HADES tracking system and the low-mass tails are partly due
to lepton energy loss by bremsstrahlung.  The overall mass resolution is comparable
to the one achieved typically with electromagnetic calorimeters.  Finally, from the
observed yields inclusive meson multiplicities can be determined by correcting
for acceptance and efficiency effects.

Note also that the $\eta \rightarrow \pi^+ \pi^- \pi^0$ and $\eta \rightarrow 3 \pi^0$
decays contribute to inclusive pion production.  The latter of these two decay modes
results in six final-state photons of which any combination of two can contribute via
conversion to the measured \eeee\ signal.  We have checked in simulations that these
correlated lepton combinations lead to a broad structure in the $M_{e^+e^-e^+e^-}$
invariant-mass distribution.
Considering that the eta is a factor 20 less abundant than the pion (see Sec.~\ref{yields})
it is not surprising that this contribution is indistinguishable from the
uncorrelated CB in Fig.~\ref{Meeee}.

\begin{table}[htb]

\caption[]{Average conversion probabilities $\langle P_{conv} \rangle$ of decay photons in
          various inner parts of the HADES setup obtained from GEANT3 simulations.
          The last row gives the cumulated probability due to all materials contributing
          to the detection of $\pi^0$ and $\eta$ mesons.  See text for a discussion of
          systematic uncertainties on those numbers.
}

\vspace*{0.2cm}
\begin{center}
\begin{tabular}{ l  c c } 
  \hline
  \hline
  Material & ~$\langle P_{conv}\rangle (\pi^0)$~ &  ~$\langle P_{conv}\rangle (\eta)$~ \\
  \hline
  target (Nb) & 2.17\% & 2.54\% \\
  target holder (C) & 0.12\% & 0.14\% \\
  beam pipe (C) & 0.46\% & 0.51\% \\
  radiator (C$_4$F$_{10}$) & 0.79\% & 0.92\% \\
  \hline
  cumulated & 3.53\% & 4.11\% \\
  \hline
  \hline

\end{tabular}
\end{center}
\label{convtable}
\end{table}

\begin{table}[hb]

\caption[]{Average effective branching ratios $\langle BR_{eeee}\rangle$, geometric acceptances
      with respect to $4\pi$ ~$\langle${\em acc}$\rangle$, pair reconstruction efficiencies
      $\langle${\em eff}$\rangle$,
      and total detection efficiencies $\langle${\em eff}$_{tot}\rangle$
      relevant for the reconstruction of the $\pi^0,\eta \rightarrow$ \eeee\ final states.
      Statistical errors due to the finite size of the simulated sample are of the order of
      3\% for $\pi^0$ and 2\% for $\eta$ on all listed quantities.  Systematic uncertainties
      are discussed in the text.
}

\vspace*{0.2cm}
\begin{center}
\begin{tabular}{ l  c c c c } 
  \hline
  \hline
  Particle & $\langle${\em BR}$_{eeee}\rangle$ & $\langle${\em acc}$\rangle$ & $\langle${\em eff}$\rangle$ & $\langle${\em eff}$_{tot}\rangle$ \\
  \hline
$\pi^0$~ & ~$1.68\times10^{-3}$ ~&~ $4.53\times10^{-3}$ ~&~ 0.063 ~&~ $4.78\times10^{-7}$\\ 
$\eta$~ & ~$9.72\times10^{-4}$ ~&~ $2.87\times10^{-2}$ ~&~ 0.11 ~&~ $3.03\times10^{-6}$\\  
  \hline
  \hline

\end{tabular}
\end{center}
\label{efftable}
\end{table}

We have performed extensive detector simulations to study the reconstruction efficiency of the 
conversion technique.  To do this we have generated meson distributions with the Pluto event
generator \cite{pluto1}, tracked the resulting particles through a realistic model of the HADES
setup with the GEANT3 physics simulation tool \cite{geant3}, embedded those tracks into real
events from the p+Nb experiment, and reconstructed the overlayed events with the full HADES
lepton analysis.  The purpose of the embedding was to include realistic detector noise into
the procedure.  As the lepton identification in HADES relies primarily on the RICH detector,
only conversion pairs produced in the inner parts of the setup can contribute to the
\eeee\ signal.  These are the niobium target segments, the carbon-composite target holder,
the carbon-composite beam pipe, and the RICH radiator gas C$_4$F$_{10}$.  Conversion pairs
produced in the RICH mirror or in any of the following materials are not detectable.
Average conversion probabilities $\langle P_{conv} \rangle$ of $\pi^0$ and $\eta$ decay
photons obtained from the simulations are listed in Table~\ref{convtable}.
More than half of the conversion takes place in the niobium targets and the rest
in the target holder, the beam pipe, and the RICH converter gas, with a the
cumulated probability of 3.5--4.1\% per photon.  The difference in probabilities
for $\pi^0$ and $\eta$, respectively, is related to the energy dependence
of the conversion process.  As the HADES pair vertex resolution, of the order
of 2--3~cm, is not good enough to cleanly isolate the various converter parts in
the event reconstruction we have refrained from applying specific vertex cuts
and have exploited the cumulated conversion effect.

Systematic errors on the total conversion probability result mainly from uncertainties
on the material budget and the alignment of the relevant inner detector parts.
The thickness of the target foils is known with an error of 2\% and their misalignment
in the beam pipe adds another 4\% error.  Inhomogeneities of the carbon-composite
material lead to an estimated uncertainty of 10\% on the holder and pipe contributions.
Finally, for the radiator contribution we assume a 5\% error.  This leads then to an overall
systematic error on the total material budget of 5\% and hence an error of about 10\% on
the effciency for detecting double conversion events.

\begin{figure}[!ht]

  \mbox{\epsfig{width=0.83\linewidth, figure=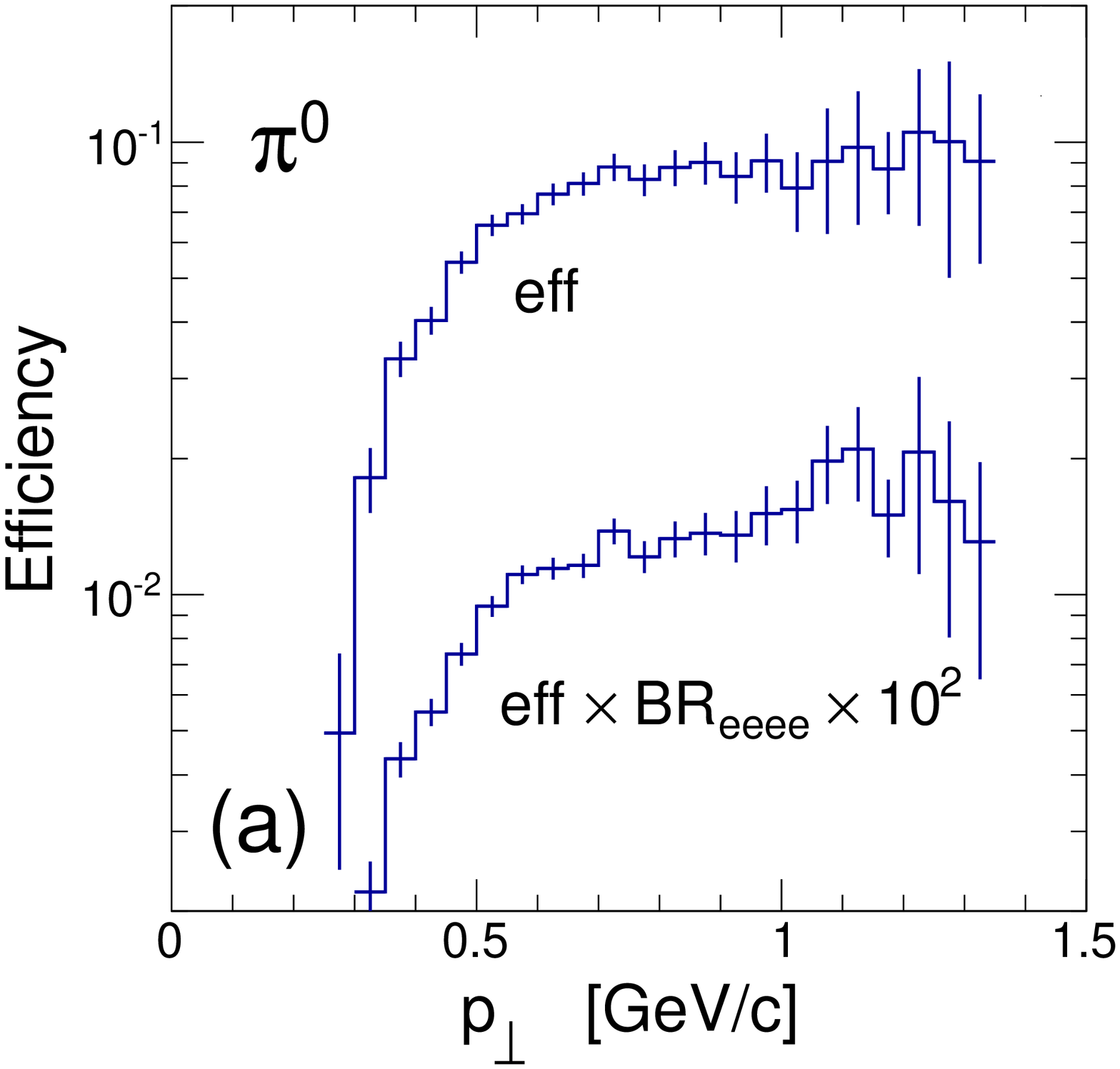}}
  \mbox{\epsfig{width=0.83\linewidth, figure=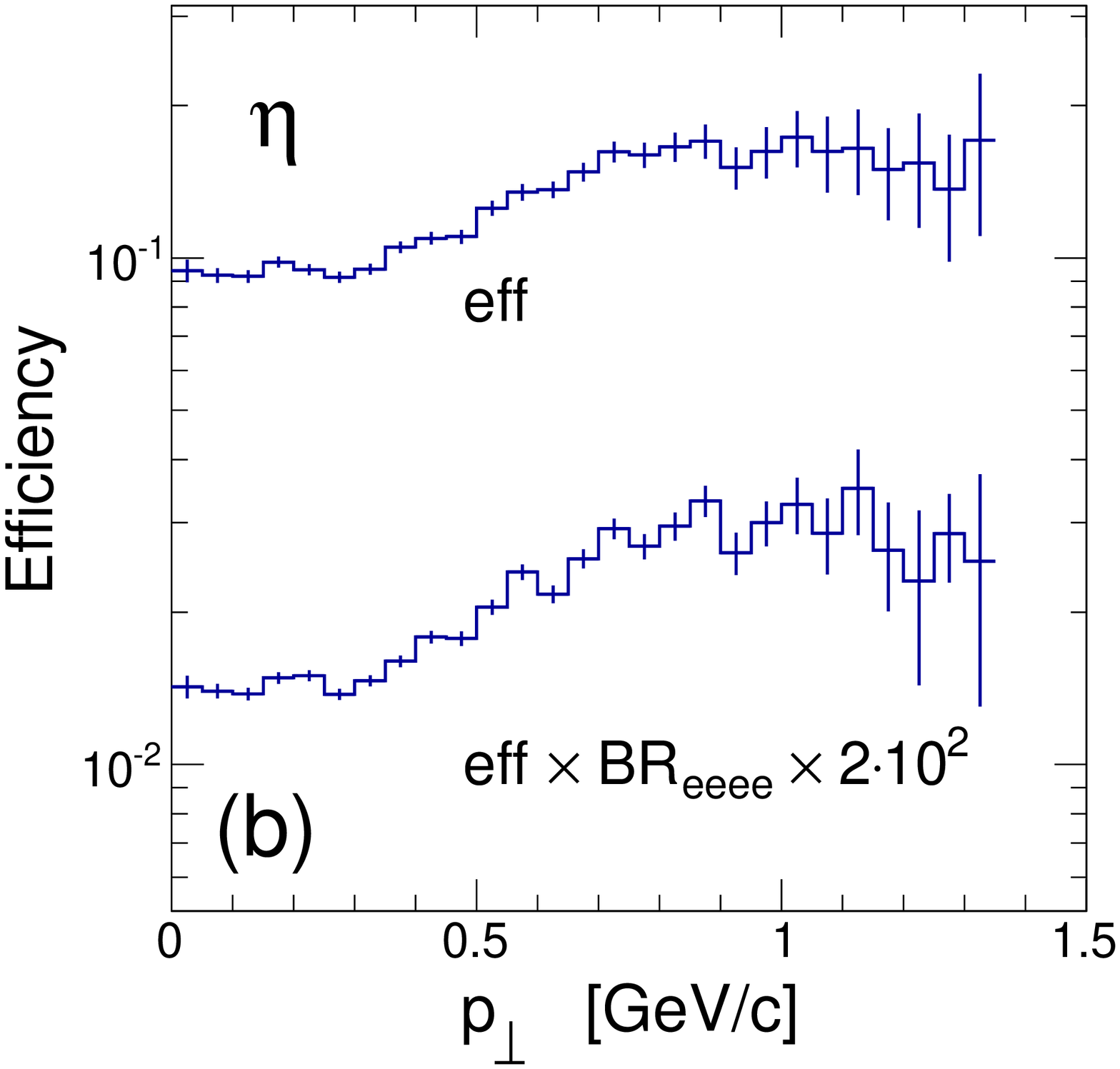}}
  \caption[]{Efficiencies of meson reconstruction from \eeee\ events
   as a function of transverse momentum $p_{\perp}$.  Shown are the $\pi^0$ (a)
   and $\eta$ (b) efficiencies with ({\em eff}~$\times BR_{eeee}$) and without ({\em eff}\,)
   the contribution of the photon conversion.  Error bars are statistical.
    }
  \label{effpt}
\end{figure}

As we do not apply a secondary vertex selection to the \ee\ pairs also Dalitz decays
are included in our pair signal.  In the Dalitz case, of course, only the one real
photon is required to convert.  The combined branching into all contributing final
states ($\gamma\gamma$, $\gamma\gamma^*$, and $\gamma^*\gamma^*$) amounts to $\simeq100$\%
for the $\pi^0$ and to $\simeq40$\% for the $\eta$ \cite{pdg2012}.

The meson sources used in the simulation were modeled according to a
fireball \cite{boltzmann,schnedermann} characterized by a temperature in the
range $T$= 80--90~\mev\ and a central laboratory rapidity in the range $y_{max}$= 0.92--0.96.
As discussed below, the choice of these values is motivated by the data itself
(cf. also Figs.~\ref{mult_pt} and \ref{mult_y}).  The Monte Carlo shows that the total detection
efficiency {\em eff}$_{tot}$ depends only weakly on the meson source properties and mostly
on the photon conversion probability, on the geometric detector acceptance with
respect to $4\pi$ {\em acc}, and on the pair reconstruction efficiency {\em eff}.
In this definition, {\em acc} includes the lepton low-momentum cutoff at around 50~\mev\
due to the track bending in the HADES magnet field and {\em eff} accounts for all detection
and reconstruction losses within the HADES acceptance.  As all of these quantities are
averaged over the two-photon, the Dalitz, and the small direct decay channels it is useful
to introduce an effective branching ratio {\em BR}$_{eeee}$ into the \eeee\ final state
which includes the photon conversion probability.  Table~\ref{efftable} summarizes the
result of our $\pi^0 \rightarrow$ \eeee\ and $\eta \rightarrow$ \eeee\ event simulations
showing in particular that the total detection efficiencies are of order $10^{-7}$ -- $10^{-6}$.
 
Systematic errors on the simulated efficiencies result from the uncertainties on the
conversion probabilities (5\% on $P_{conv}$, 7.5\% on $BR_{eeee}$), on the branching ratios
of the contributing decays (1\% on $BR_{eeee}$ for $\pi^0$ and 2\% for $\eta$, from \cite{pdg2012}),
and on the detector and reconstruction efficiencies (10\%, from a comparison of various
simulated and measured observables in the HADES detector \cite{hades_tech}).  Combining
all of these contributions we assign to the total efficiency a conservative systematic
error of 15\%.

The transverse momentum ($p_{\perp}$) dependence of the meson reconstruction efficiency
is depicted in Fig.~\ref{effpt}, with and without the photon conversion probability included.
The cutoff of the $\pi^0$ efficiency at $p_{\perp}\lesssim0.35$~\gevc\ is caused mostly by the
strong bending of low-momentum tracks, i.e. those with $p<0.1$~\gevc, in the HADES magnetic
field.  Because of the large mass of the eta meson, its efficiency is much less afflicted
by low-momentum tracks and consequently reconstruction is possible down to zero~$p_{\perp}$.

\section{Meson yields}
\label{yields}

\subsection{Negative pions}

\begin{figure}[!ht]

  \mbox{\epsfig{width=0.99\linewidth, figure=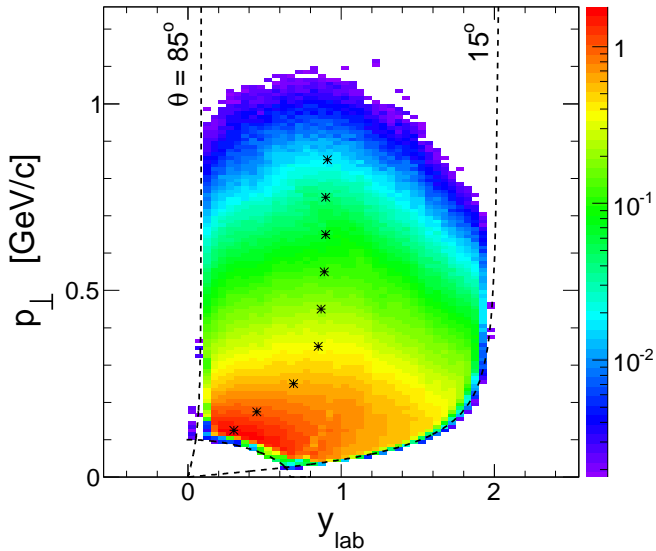}}
  \vspace*{-0.2cm}
  \caption[]{(Color online) Efficiency-corrected $d^2N/dy\,dp_{\perp}$ distribution
   of negative pions detected by HADES in the 3.5~\gev\ p+Nb reaction.
   The color scale indicates yield per unit rapidity, per \gev, and per minimum-bias event.
   The dashed lines delineate the geometric acceptance (including a $p>0.12$~\gevc\ cut),
   and the stars indicate loci of the maximum of the $dN/dy$ distribution, $y_{max}$, as function of
   transverse momentum.
   }
  \vspace*{-0.2cm}
  \label{hades_pim}
\end{figure}

As pointed out above, besides measuring e$^+$ and e$^-$, the HADES detector also provides
high-quality data on charged hadrons.  Here we use the concurrently measured
negative pions \cite{hades_pim} to validate the reconstructed $\pi^0$ yields and
extrapolate them to $p_{\perp}<0.35$~\gevc.  Figure~\ref{hades_pim} shows
efficiency-corrected\footnote{The $\pi^-$ yields are corrected as well for the gaps caused in the azimuthal
acceptance by the HADES magnet coils.} double differential $\pi^-$ yields $d^2N/dy\,dp_{\perp}$ adapted
from \cite{hades_pim}.  The charged pion acceptance is constrained geometrically
to polar angles of $15^{\circ}$ -- $85^{\circ}$ as well as by a momentum cut of
$p>0.12$~\gevc.  This results in an acceptance in laboratory rapidity of $y_{lab}\simeq$~0.2 -- 1.8
and in transverse momentum of $p_{\perp}\gtrsim0.1$~\gevc.  Because of this quite large
coverage one can expect that the extrapolation to full phase space will lead
to moderate systematic uncertainties only.  At a bombarding energy of 3.5~\gev\
the rapidity of the nucleon-nucleon center-of-mass system is at $y_{NN} = 1.12$.
From the Fig.~\ref{hades_pim} it is however apparent that in the p+Nb reaction the pion
yield is not peaked at a mid-rapidity $y_{NN}$ but at a lower value with, in addition,
a marked $p_{\perp}$ dependence.  Fits of a Gaussian function to $dN/dy$ projections
done for various $p_{\perp}$ slices give the loci $y_{max}$ of the maximum yield vs. $p_{\perp}$,
as shown by stars in Fig.~\ref{hades_pim}.  In particular, low-$p_{\perp}$ pions seem
to be radiated mostly from a target-like source, near $y = 0$, pointing to a high degree
of stopping of the incoming projectile.  Obviously not only first-chance nucleon-nucleon
collisions contribute to pion emission in the p+Nb reaction, but proton elastic and inelastic
rescattering followed by secondary production processes add a soft target-like component.
This is also corroborated by various transport-model calculations \cite{hsd2,urqmd2,gibuu2}.

Integrating $d^2N/dy\,dp_{\perp}$ within the HADES rapidity coverage we find
an accepted $\pi^-$ yield of 0.50~per LVL1 event.
From simulations we know
that the LVL1 trigger leads to a 42\% enhancement of the average detected
charged-pion yield per event (see \cite{hades_pim}).  Correcting for this trigger bias
we obtain an accepted yield of 0.35 per p+Nb reaction.  The Gauss fits done to the $dN/dy$
projections provide furthermore a means to extrapolate the measured $\pi^-$ yield outside
of the HADES rapidity coverage.  Alternatively, transport models, e.g. HSD \cite{hsd1},
UrQMD \cite{urqmd2}, or GiBUU \cite{gibuu2} (see Sec.~\ref{transport} below), can be used to perform
the extrapolation in $y$ and $p_{\perp}$ to full solid angle.  In fact, integrating the yield within
the geometric acceptance limits, extrapolating it either way --- via Gauss fits to $dN/dy$
projections in $p_{\perp}$ slices or, better, with the help of transport calculations done
for p+Nb --- and correcting for the LVL1 bias we obtain on average a minimum-bias
inclusive $\pi^-$ multiplicity of $N_{\pi^-}=0.60$.  The error on this corrected $\pi^-$ multiplicity
is dominated by systematic effects introduced mostly by the correction of the LVL1 trigger
bias ($\pm13$\%) and the spread in the model-dependent extrapolation in phase space (-10\%, +15\%);
statistical errors are however negligible.  Table~\ref{multtable} lists the extracted
multiplicity values and their associated uncertainties.

\begin{table*}[!hbt]

\caption[]{Integrated minimum-bias inclusive meson multiplicities per p+Nb collision
           $N_{acc}$, within the accepted rapidity range $0.2<y_{lab}<1.8$, and
           $N_{4\pi}$, extrapolated to full solid angle.  Statistical and systematic
           uncertainties are given; statistical errors are negligibly small for $\pi^-$.
}

\vspace*{0.2cm}
\begin{center}
\begin{tabular}{ l  c c } 
  \hline
  \hline
  Particle & $N_{acc}$ & $N_{4\pi}$ \\
  \hline
\rule{0pt}{3ex}$\pi^-$~&~ $0.35\pm0.05 \;(sys) $ ~&~ $0.60\pm0.10 \;(sys)$ \\ 
\rule{0pt}{3ex}$\pi^0$~&~ $0.39\pm0.06 \;(stat) \pm0.08 \;(sys) $ ~&~ $0.66\pm0.09 \;(stat)\; \pm0.17 \;(sys)$ \\
\rule{0pt}{3ex}$\eta$~&~ $0.031\pm0.002 \; (stat) \pm0.007 \;(sys) $ ~&~ $0.034\pm0.002 \;(stat)\; \pm0.008 \;(sys)$ \\[1.5mm]
  \hline
  \hline

\end{tabular}
\end{center}
\label{multtable}
\end{table*}

\subsection{Neutral mesons}

We come now to the presentation of our differential $\pi^0$ and $\eta$ yields.
First notice that the LVL1 trigger (meaning at least three charged hits in the TOF wall)
does not introduce a bias on events with an \eeee\ signature.  Likewise,
the LVL2 trigger efficiency is found to be \mbox{$>99\%$} for such events.
Hence no explicit corrections for trigger effects are needed.  The systematic
uncertainties on the \eeee\ observables to be taken into account are those introduced by
the efficiency correction ($\pm15$\%), the CB subtraction ($\pi^0$: $\pm5$\%, $\eta$: $\pm10$\%),
and the model-dependent extrapolation to full solid angle ($\pi^0$: $\pm15$\%, $\eta$: $\pm10$\%).

\begin{figure}[!htb]

  \mbox{\epsfig{width=0.99\linewidth, height=0.99\linewidth, figure=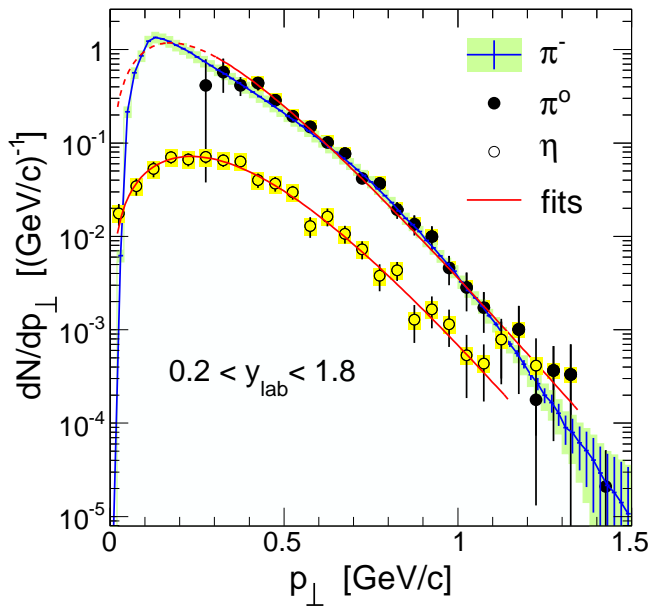}}
  \vspace*{-0.2cm}
  \caption[]{(Color online) $\pi^0$ (full circles), $\pi^-$ (blue histogram), and $\eta$ (open circles)
    transverse momentum distributions $dN/dp_{\perp}$ per minimum-bias event
    in 3.5~\gev\ p+Nb reactions within the HADES rapidity and momentum acceptance.
    The latter leads to a $p_{\perp}\gtrsim0.35$~\gevc\ cut for the $\pi^0$. 
    Statistical errors are shown as vertical bars, systematic errors as yellow and green shaded boxes.
    The red solid curves are Boltzmann fits to the $\pi^0$ and $\eta$ data (see text for details).
  }

  \vspace*{-0.2cm}
  \label{mult_pt}
\end{figure}

\begin{figure}[!ht]

  \mbox{\epsfig{width=0.80\linewidth, height=0.80\linewidth, figure=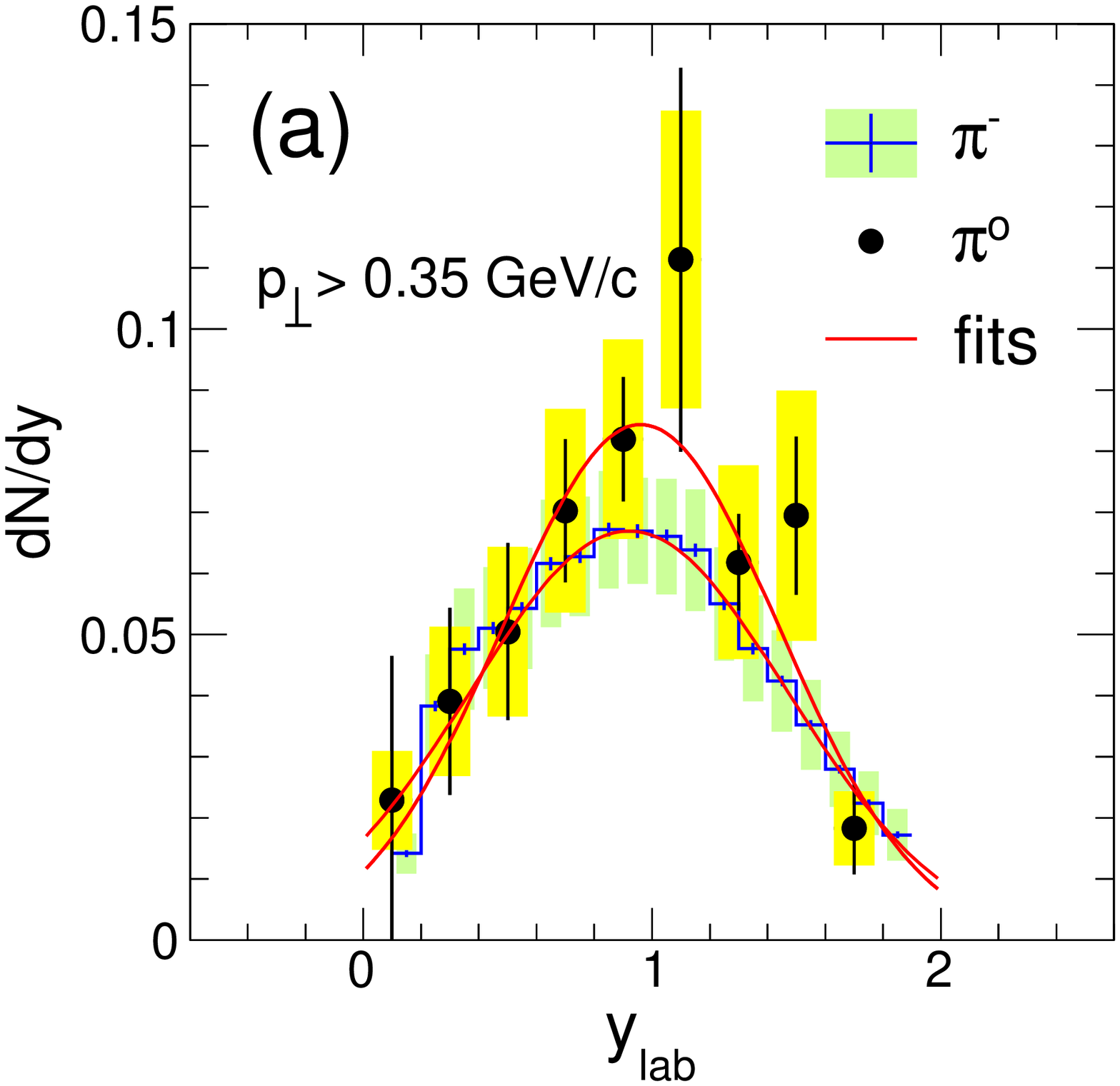}}
  \vspace*{-0.2cm}
  \mbox{\epsfig{width=0.80\linewidth, height=0.80\linewidth, figure=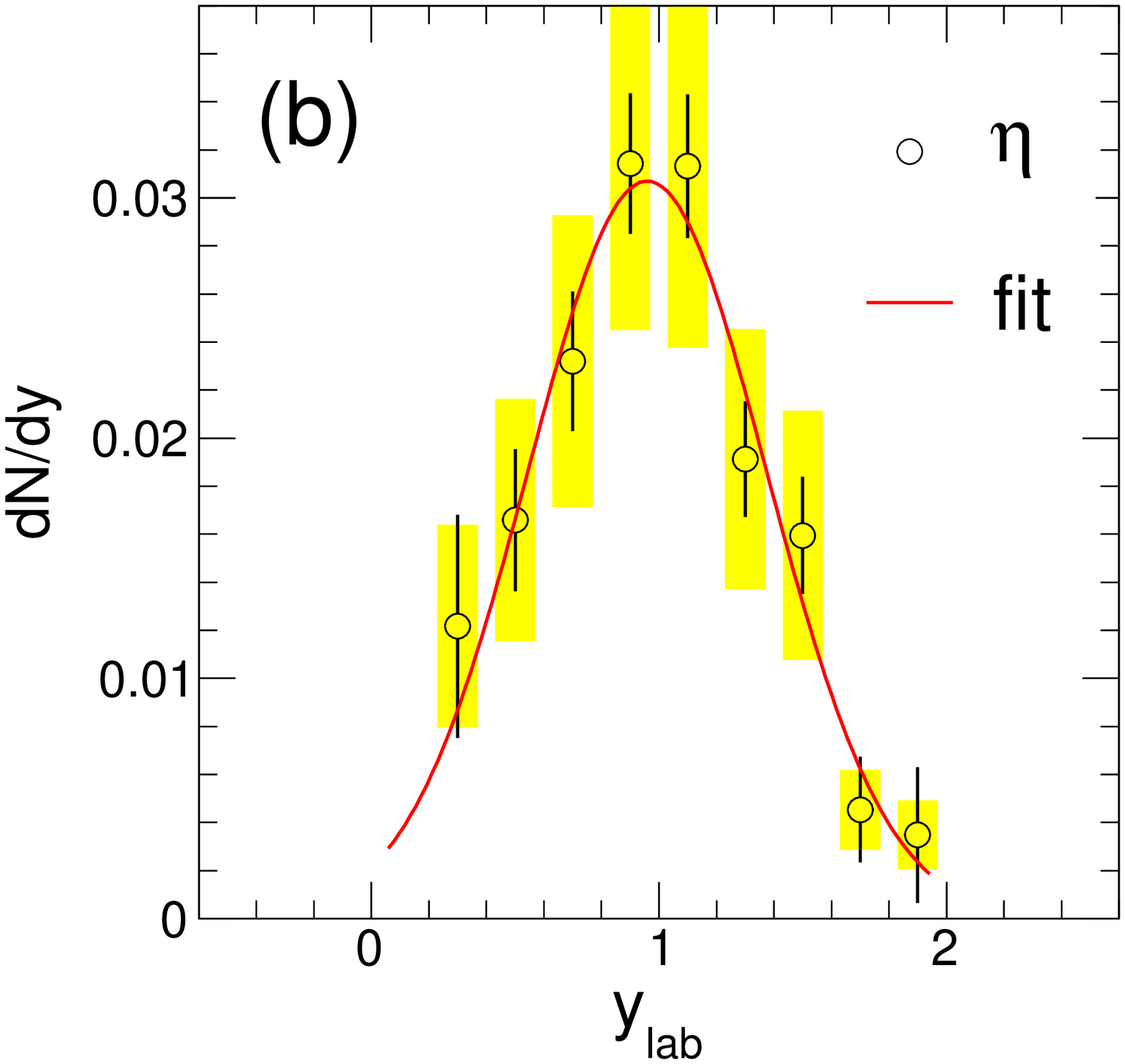}}
  \vspace*{-0.2cm}
  \caption[]{(Color online) (a) $\pi^0$ (negative pion yield shown as histogram)
    and (b) $\eta$ rapidity distributions $dN/dy$ per minimum-bias event
    in 3.5~\gev\ p+Nb reactions.  Pions are shown for $p_{\perp}>0.35$~\gevc, $\eta$ in full
    $p_{\perp}$ range.
    Statistical errors are indicated by vertical bars, systematic errors are depicted
    by yellow shaded boxes.  Red solid curves are Gauss fits to the data
    ($\pi^0$: $y_{max}=0.94\pm0.05$, $\sigma_y=0.48\pm0.06$, $\chi^2/df=6.1/6$;
     $\pi^-$: $y_{max}=0.91\pm0.01$, $\sigma_y=0.55\pm0.02$, $\chi^2/df=17.2/16$;
     $\eta$: $y_{max}=0.96\pm0.03$, $\sigma_y=0.41\pm0.03$, $\chi^2/df=5.2/6$).
  }

  \vspace*{-0.2cm}
  \label{mult_y}
\end{figure}

As is visible from the invariant-mass spectrum in Fig.~\ref{Meeee}, both neutral mesons
can be selected with appropriate mass cuts.  We have used $0.10 < M < 0.16$ for the $\pi^0$
and $0.46 < M < 0.60$ for the $\eta$.  Subtracting the corresponding
combinatorial background and applying corrections for photon conversion, as well as
for lepton track identification and reconstruction efficiencies the double-differential
yields $d^2N/dydp_{\perp}$ are obtained as a function of rapidity and transverse momentum.
As stated in Sec.~\ref{method}, our efficiency corrections are based on the reconstruction
of simulated meson decays embedded into real events.  To do a first correction,
we started out with relativistic Boltzmann distributions of an assumed temperature of 100~\mev\
and then refined this value in a second pass.  The same holds for the central rapidities
of the $\pi^0$ and $\eta$ sources which, like in case of the $\pi^-$, are observed to be
substantially below the $y_{NN}=1.12$ value.  All corrections were furthermore done
concurrently in two dimensions, $y$ and $p_{\perp}$, in order to alleviate any remaining
dependence on our assumptions about the meson source characteristics.  The resulting
final $dN/dp_{\perp}$ and $dN/dy$ distributions, normalized per minimum-bias event,
are displayed in Figs.~\ref{mult_pt} and \ref{mult_y}.  For comparison, the negative
pion distributions are also shown.

\begin{figure}[!hbt]
  \mbox{\epsfig{width=0.99\linewidth, figure=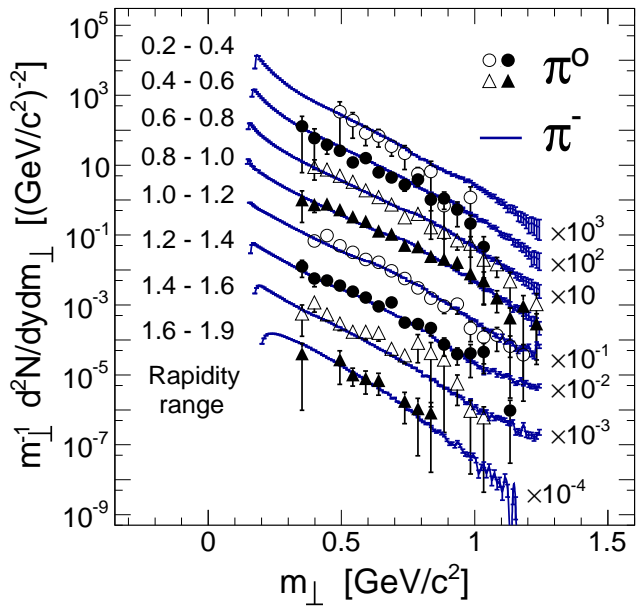}}
  \vspace*{-0.2cm}
  \caption[]{Transverse mass distributions per minimum-bias event $m^{-1}_{\perp} dN/dm_{\perp}$
    of measured $\pi^-$ and $\pi^0$ for the listed rapidity cuts.
    Error bars are statistical; systematic errors (not shown) on the 
    $\pi^-$ ($\pi^0$) yields are $\pm13$\% ($\pm19$\%).
    }
  \vspace*{-0.2cm}
  \label{pion_Mt}
\end{figure}

Due to the efficiency cutoff discussed in Sec.~\ref{method}, no $\pi^0$ yield is detected
at low $p_{\perp}$.  This is directly visible in Fig.~\ref{mult_pt} and it is also the
reason why in Fig.~\ref{mult_y} we show both pion rapidity distributions (i.e. $\pi^0$
and $\pi^-$) for $p_{\perp}$ larger than 0.35~\gevc\ only.  For the $\eta$ meson, however,
the $p_{\perp}$ coverage is complete.  The rapidity coverage of both neutral mesons
is restricted to $y_{lab}$ = 0.2 -- 1.8 by the detector geometry.  In a first attempt
to characterize the observed meson yields we confront them with the isotropic
fireball model \cite{boltzmann,schnedermann}.  Adjusting a Boltzmann type distribution
$dN/dp_{\perp} \propto p_{\perp} m_{\perp} K_1(m_{\perp}/T)$ to the meson transverse momentum
distribution and a Gaussian $dN/dy \propto \exp{(-0.5(y-y_{max})^2/\sigma_y^2)}$ to the
rapidity distribution we find $T=92\pm3$~\mev\ ($\chi^2/df=17.4/18$),
$y_{max}=0.94\pm0.05$, and $\sigma_y=0.48\pm0.06$ ($\chi^2/df=6.1/6$)
for the $\pi^0$, respectively $T=84\pm3$~\mev\ ($\chi^2/df=14.1/21$),
$y_{max}=0.96\pm0.03$, and $\sigma_y=0.41\pm0.03$ ($\chi^2/df=5.2/6$)
for the $\eta$.  Here $K_1(x)$ is the modified Bessel function, $m_{\perp} = \sqrt{p_{\perp}^2 + m^2}$
is the transverse mass, $T$ is the fitted temperature parameter, $y_{max}$ is the average source
mid-rapidity (actually the peak position thereof), and $\sigma_y$ is the width of the
accepted rapidity distribution.  The fit results show in particular that, within error bars,
the rapidity distributions of both pion species agree in shape.

To characterize the pion source further, Fig.~\ref{pion_Mt} shows the pion transverse-mass distributions
$m^{-1}_{\perp} \; dN/dm_{\perp}$ projected for various rapidity selections.  It is apparent from this figure,
and also Fig.~\ref{mult_pt}, that a Boltzmann source does not describe very well the low-$p_{\perp}$ behavior.
On the other hand, both pion species display in general a very similar behavior as function of $m_{\perp}$ and $y$.
Only in the first rapidity bin ($y_{lab}$=0.2 -- 0.4) the $\pi^-$ show a somewhat harder spectrum
than the $\pi^0$, which we attribute to a contamination of the $\pi^-$ spectrum at small polar angles
with fake tracks.  We prefer, still, to use directly the shape of the measured negative pion
distribution\footnote{ To do this the $\pi^-$ spectrum was first corrected for its Coulomb shift
$E_C = -1.44 (Z_{Nb}+1)/R_{Nb} = -12$~\mev, where $Z_{Nb}=41$ and $R_{Nb}=5.2$~fm.}
for extrapolating the $\pi^0$ yield below 0.35~\gevc.  Doing this, we get a yield per minimum-bias
event of $N_{\pi^0}=0.39$ within the accepted rapidity range of $0.2<y_{lab}<1.8$.
In a second step the extrapolation to full solid angle can be done, based on transport-model
calculations as discussed above, giving a minimum-bias inclusive multiplicity of $N_{\pi^0}=0.66$.
Statistical and systematic error bars on those results are given in Table~\ref{multtable}.

\begin{figure}[!hbt]
  \mbox{\epsfig{width=0.99\linewidth, figure=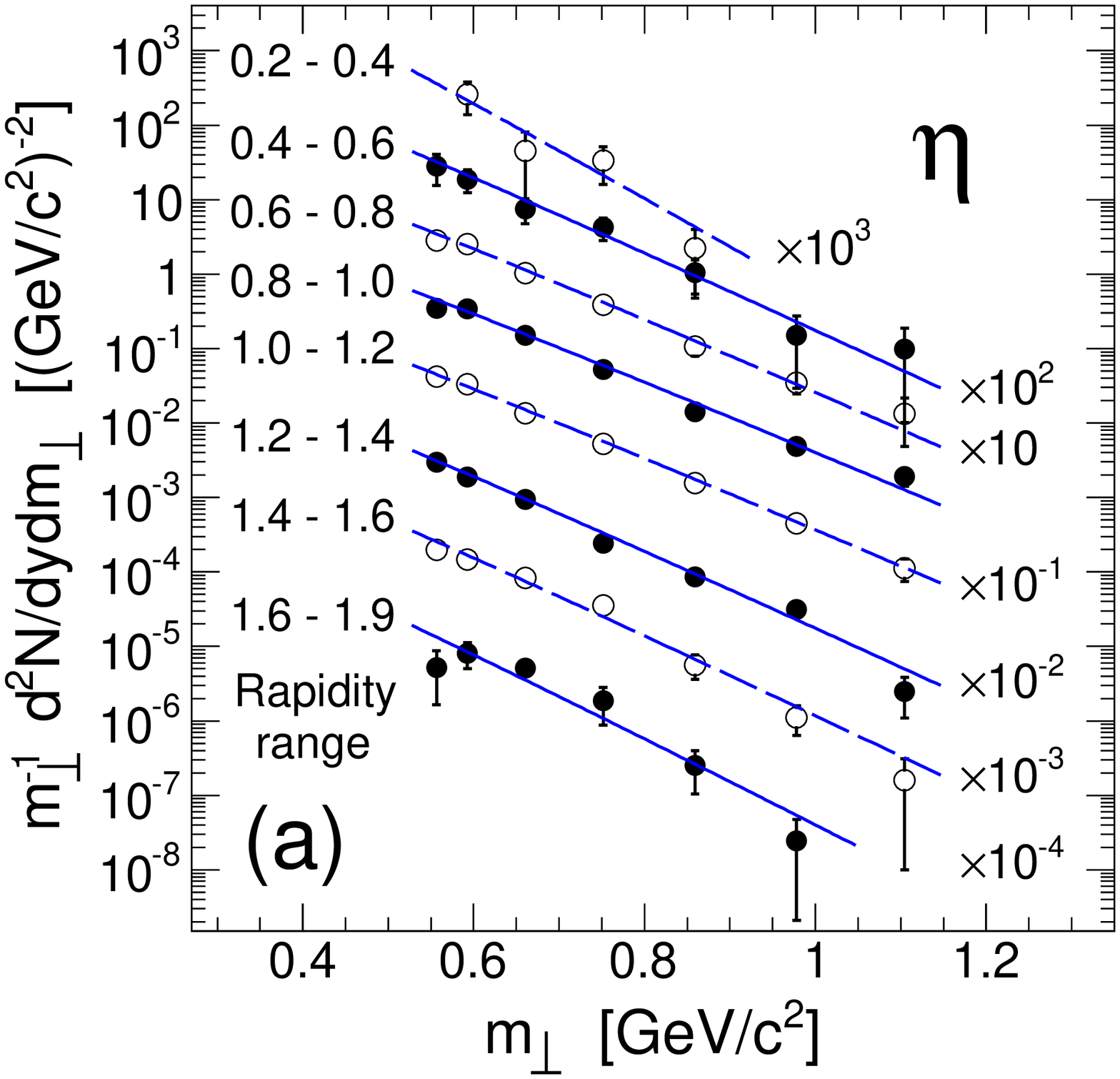}}
  \vspace*{-0.2cm}
  \mbox{\epsfig{width=0.90\linewidth, figure=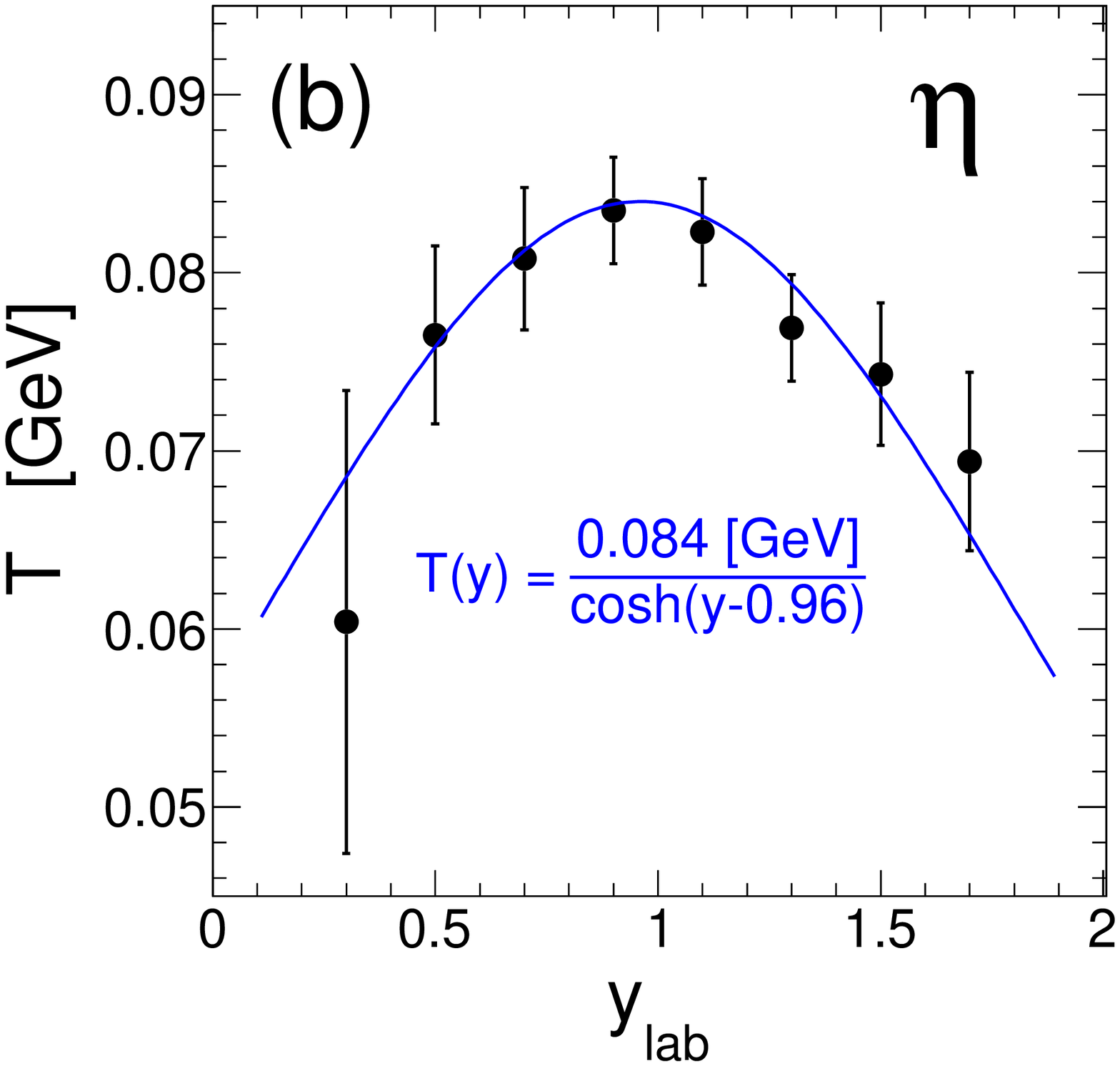}}
  \vspace*{-0.2cm}
  \caption[]{(a) Transverse mass distributions per minimum-bias event $m^{-1}_{\perp} dN/dm_{\perp}$
    of reconstructed eta mesons for the listed rapidity cuts.  The data points
    have been corrected for their shift within the rather large $m_{\perp}$ bins. 
    Error bars shown are statistical only;  systematic errors (not shown) on the 
    yields are 21\%.  The curves are exponential fits to the data.
    (b) Resulting slope parameters $T(y)$ shown as function of the rapidity.
    The solid curve corresponds to a thermal source
    of temperature $T=84$~\mev\ and central rapidity $y_{max}=0.96$.
    }
  \vspace*{-0.2cm}
  \label{eta_Mt}
\end{figure}

For the eta meson, being four times heavier than the pion, the HADES detector provides complete
transverse-momentum coverage.  The rapidity coverage, although restricted to $y_{lab}$ = 0.2 -- 1.8,
is very large too.  Figures \ref{mult_pt} and \ref{mult_y} show that
the $\eta$ phase space distribution is well described by a Boltzmann fit
in transverse momentum and by a Gaussian in rapidity.  The latter fit yields a width
$\sigma_y$, which can be related \cite{schnedermann} to the longitudinal temperature parameter
$T_{\parallel}$ of the eta source via the relation $\sigma_y = \sqrt{T_{\parallel}/M_{\eta}}$.
From $\sigma_y = 0.41\pm0.03$ and $M_{\eta} = 0.548$ one obtains $T_{\parallel}=92\pm13$~\mev\
which is within error bars still consistent with the transverse temperature parameter obtained
from the above Boltzmann fit, namely $T=T_{\perp}=84$~\mev.  Finally, Fig.~\ref{eta_Mt} shows
Boltzmann fits to the $\eta$ transverse mass distributions $m^{-1}_{\perp} \; dN/dm_{\perp}$ for
various rapidity selections, as well as the evolution with rapidity of the fitted slope parameter, $T(y)$.
All of those are compatible with the assumption of an isotropic fireball:  the $m_{\perp}$
distributions are thermal with their slope varying like $T_{\perp}/\cosh{(y-y_{max})}$
where $T_{\perp}$ is taken from the previous Boltzmann fit to $dN/dp_{\perp}$ and  
$y_{max}$ is the central rapidity obtained in the above Gauss fit to $dN/dy$. 

Integrating either of the $dN/dy$ or $dN/dp_{\perp}$ distributions we obtain a yield
of 0.031 accepted eta per reaction and, extrapolating with the help of
our fireball fits to full solid angle, an inclusive multiplicity of $N_{\eta}=0.034$.
Extrapolations based on transport models yield slightly larger values (see Sec.~\ref{transport}).
This is taken into account in the systematic uncertainties listed in Table~\ref{multtable}.

\begin{figure}[!hbt]
  \mbox{\epsfig{width=0.99\linewidth, figure=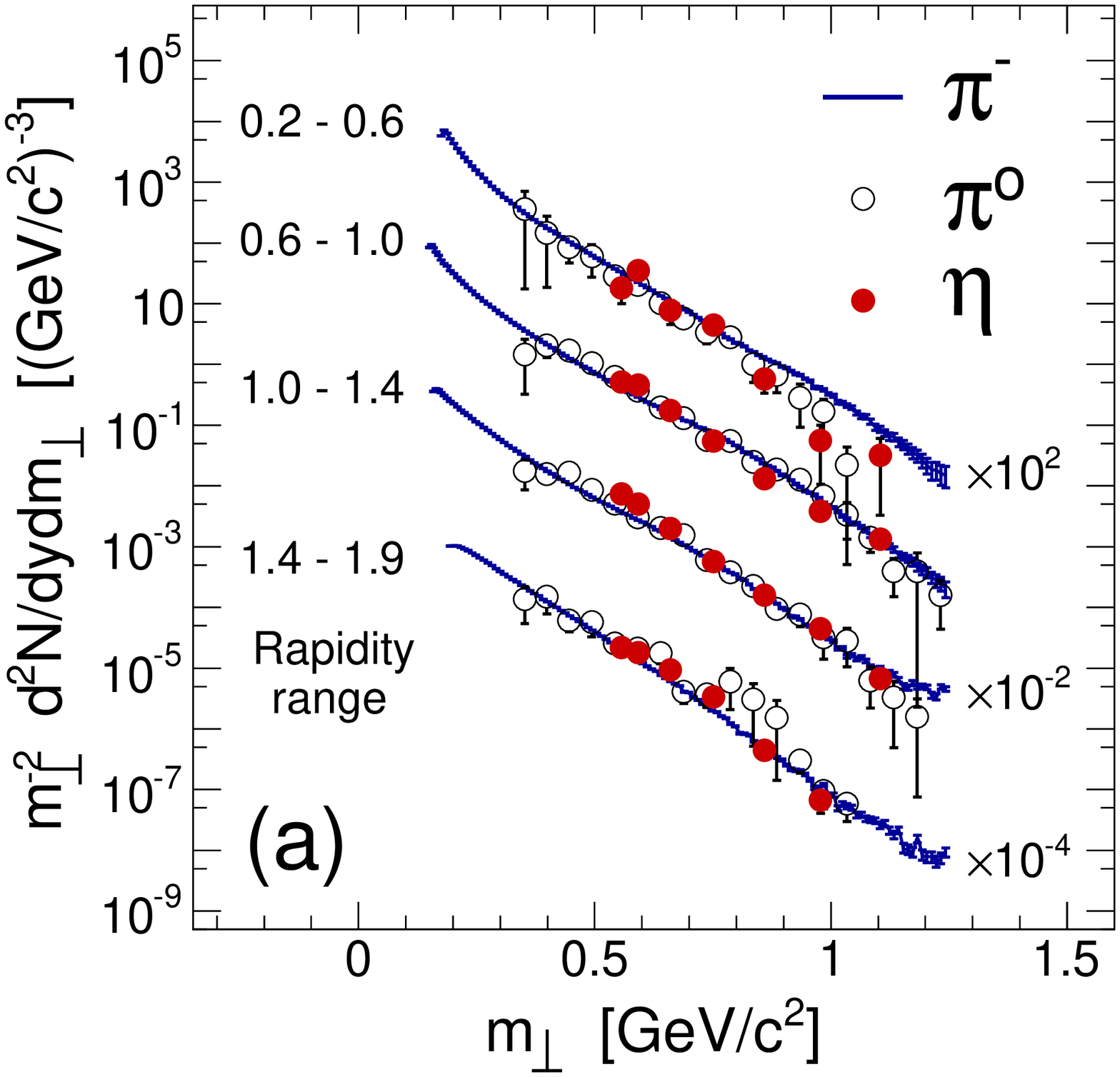}}
  \vspace*{-0.2cm}
  \mbox{\epsfig{width=0.99\linewidth, figure=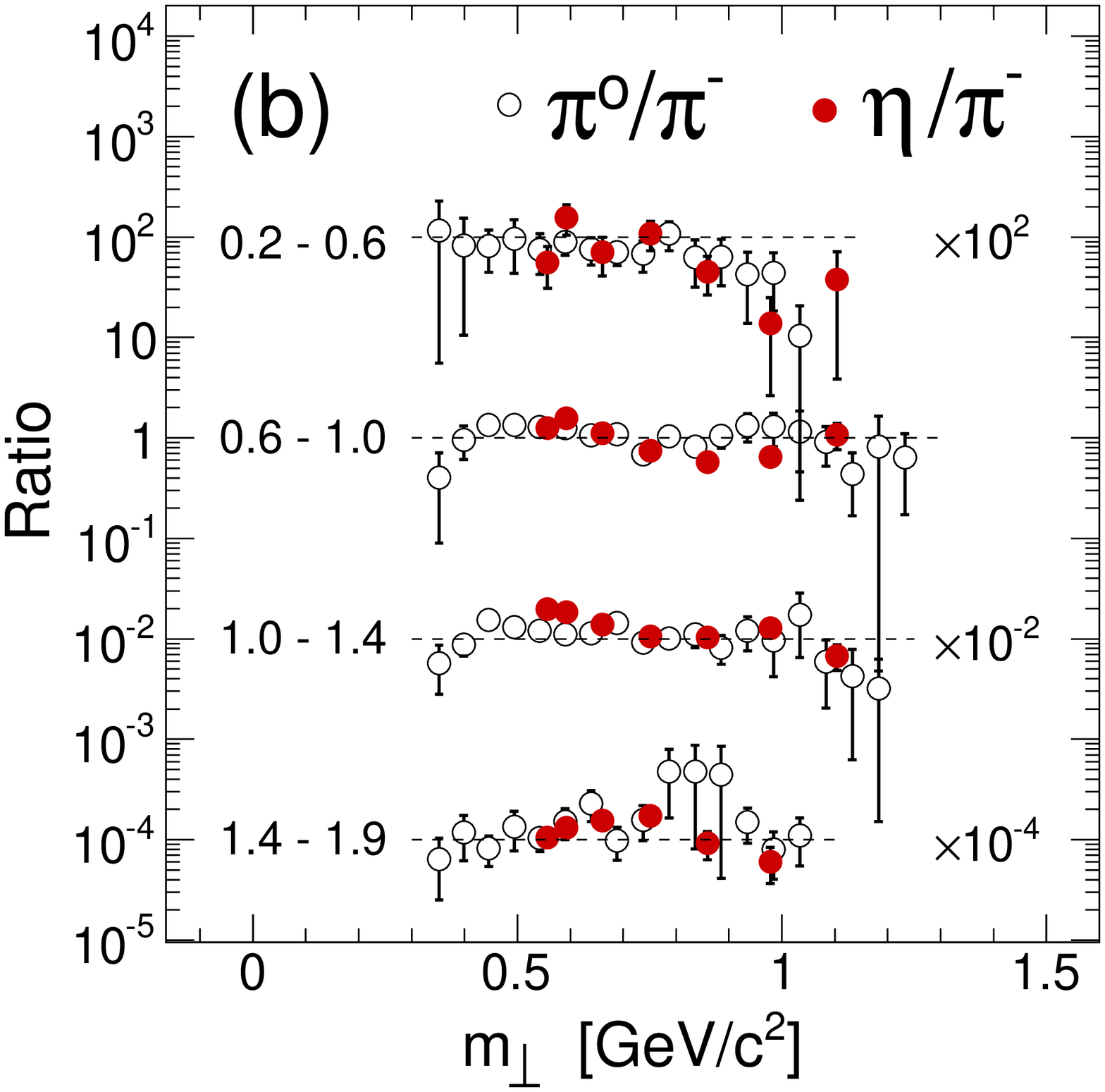}}
  \vspace*{-0.2cm}
  \caption[]{(Color online) (a) Illustration of meson $m_{\perp}$-scaling in 3.5 \gev\ p+Nb reactions
    by superimposing the reconstructed pion and eta transverse mass distributions, 
    $m^{-2}_{\perp} dN/dm_{\perp}$.  A bin-shift correction has been applied to the $\eta$ data points. 
    Four rapidity selections are shown;  error bars are statistical only.
    (b) Yield ratios, $\pi^0$ over $\pi^-$ and $\eta$ over $\pi^-$,
    as function of $m_{\perp}$ for the same rapidity selections.  
    }
  \vspace*{-0.2cm}
  \label{mtscaling}
\end{figure}

Having available differential yields of both pions and etas from the same reaction, we can compare
their scaling with transverse mass.  So-called $m_{\perp}$-scaling has indeed been found previously
for $\pi^0$ and $\eta$ production in 1~and 1.5~\gevu\ Ar+Ca collisions \cite{TAPSmtscaling}.
The observation was that the production cross sections at mid-rapidity of different mesons are
identical at a given $m_{\perp}$ value.  Model calculations have been able to reproduce
this phenomenon  \cite{QGSMmtscaling,BUUmtscaling}.  According to \cite{QGSMmtscaling}, in particular,
proton-nucleus collisions should display $m_{\perp}$-scaling as well.  In Fig.~\ref{mtscaling}(a) we
show, therefore, an overlay of our pion and eta $m_{\perp}$~distributions for a few broad rapidity bins.
We have chosen here the $m^{-2}_{\perp} \; dN/dm_{\perp}$ representation, (i) because this form
represents a Boltzmann source as approximate exponential function, and (ii) because it is particularly
well suited for visualizing $m_{\perp}$-scaling.  Despite slight differences in slope --- consistent
with the temperature parameters obtained from the Boltzmann fits discussed above --- an overall
good agreement of the $\pi^0$ and $\eta$ yields at $m_{\perp}>M_{\eta}$ is apparent.  The $\pi^-$ yields
follow the $m_{\perp}$-scaling as well, except at low rapidities.  As we discussed already in the context of
Fig.~\ref{pion_Mt}, we attribute this deviation to a contamination of fake tracks in the $\pi^-$ spectrum.
The same trends are visible in panel (b) of Fig.~\ref{mtscaling} which shows ratios of the meson yields,
namely $\pi^0/\pi^-$ and $\eta/\pi^-$.  Except for part of the low-rapidity bin, the ratios are compatible
with unity at transverse masses above $M_{\eta}$.  We conclude that, albeit the pion spectra do not
follow Boltzmann distributions at low transverse momentum, $m_{\perp}$-scaling seems to hold.
This finding suggests that the meson yields are determined mostly by phase space:
Although we can assume that meson production is mediated mostly by baryon resonance excitation, it does
not matter whether one produces an eta meson at low momentum or a pion at high momentum, as long as their
transverse mass is the same.

Finally, we want to point out that the extrapolated meson multiplicities can be
transformed into a production cross section by multiplication with the total
reaction cross section, $\sigma_{reac}$.  Parameterizations of the proton-nucleus
absorption cross section as function of bombarding energy do
exist \cite{sihver,tripathi,wellisch} and they suggest for p(3.5~\gev)+Nb
values of $\sigma_{reac}$ ranging from 990~mb (ref.~\cite{tripathi}) to 1060~mb (ref.~\cite{wellisch}).
The comparison of $\pi^-$ multiplicities measured with HADES and
interpolated HARP $\pi^-$ cross sections \cite{harp}
yields a compatible value of $\sigma_{reac}=848\pm126$~mb \cite{hades_pim}.

\section{Comparison with transport models}
\label{transport}

\begin{figure*}[!hbt]

  \mbox{\epsfig{width=0.49\linewidth, height=0.49\linewidth, figure=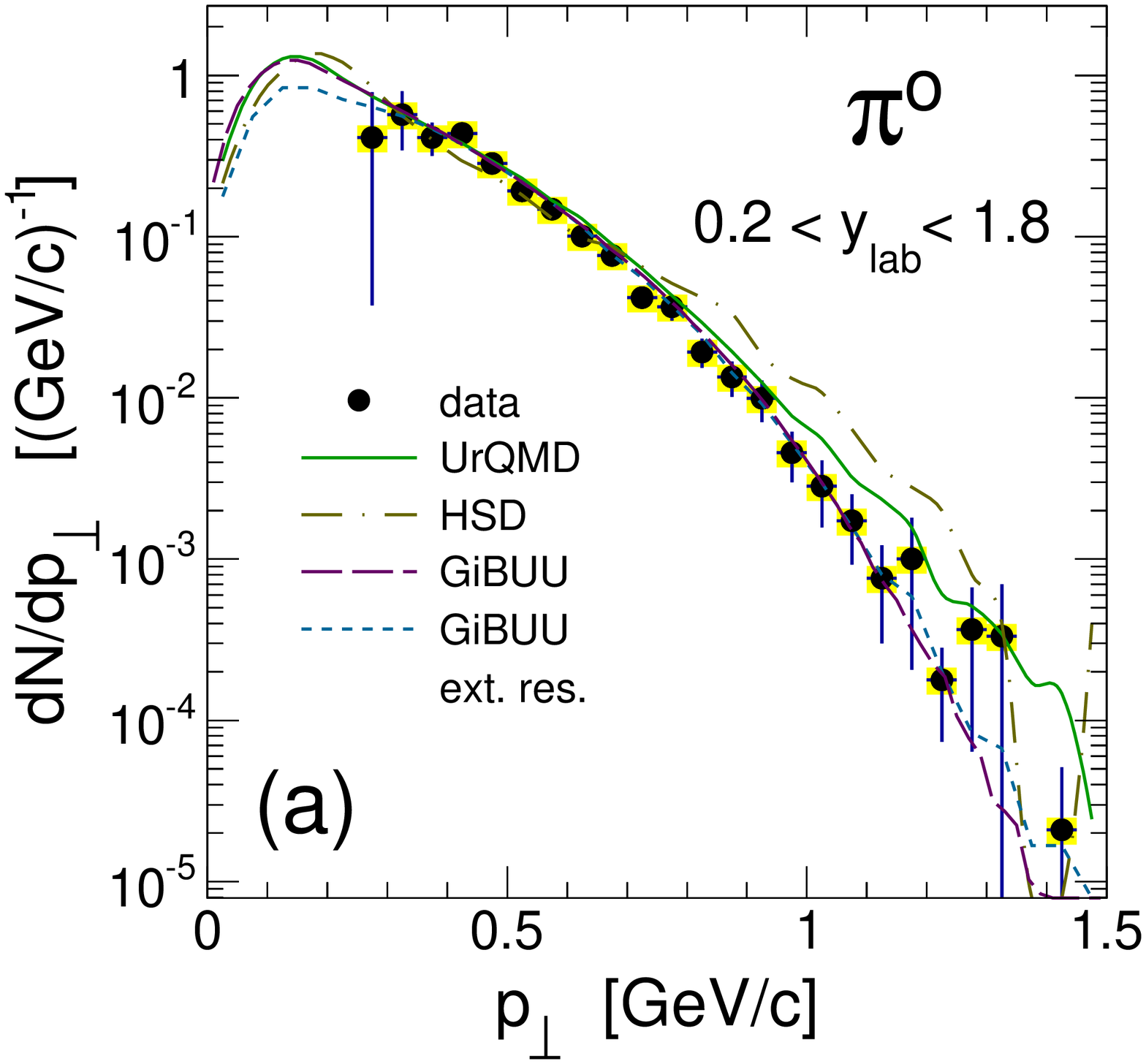}}
  \vspace*{-0.2cm}
  \mbox{\epsfig{width=0.49\linewidth, height=0.49\linewidth, figure=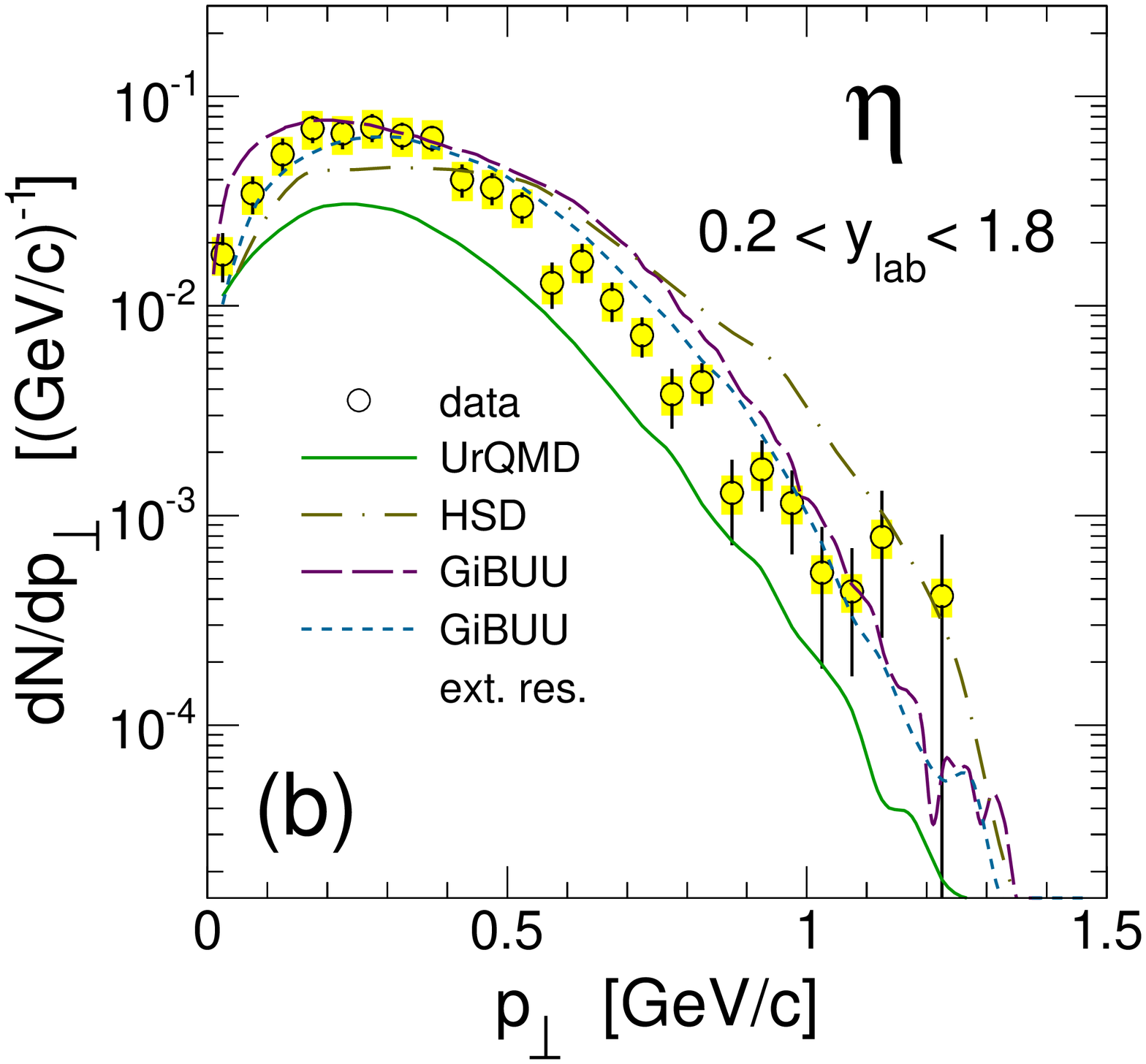}}
  \vspace*{-0.2cm}
  \mbox{\epsfig{width=0.49\linewidth, height=0.49\linewidth, figure=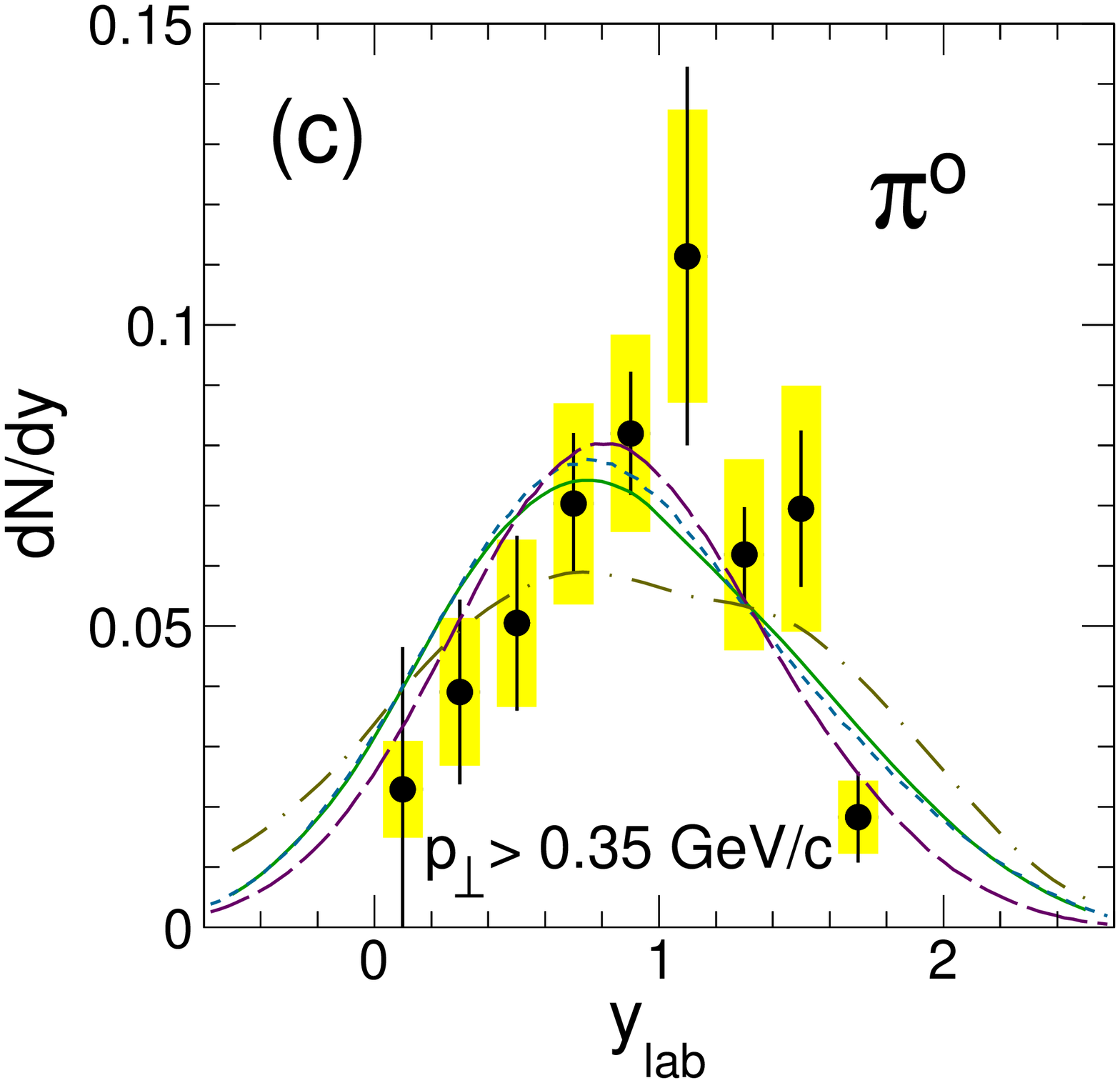}}
  \vspace*{-0.2cm}
  \mbox{\epsfig{width=0.49\linewidth, height=0.49\linewidth, figure=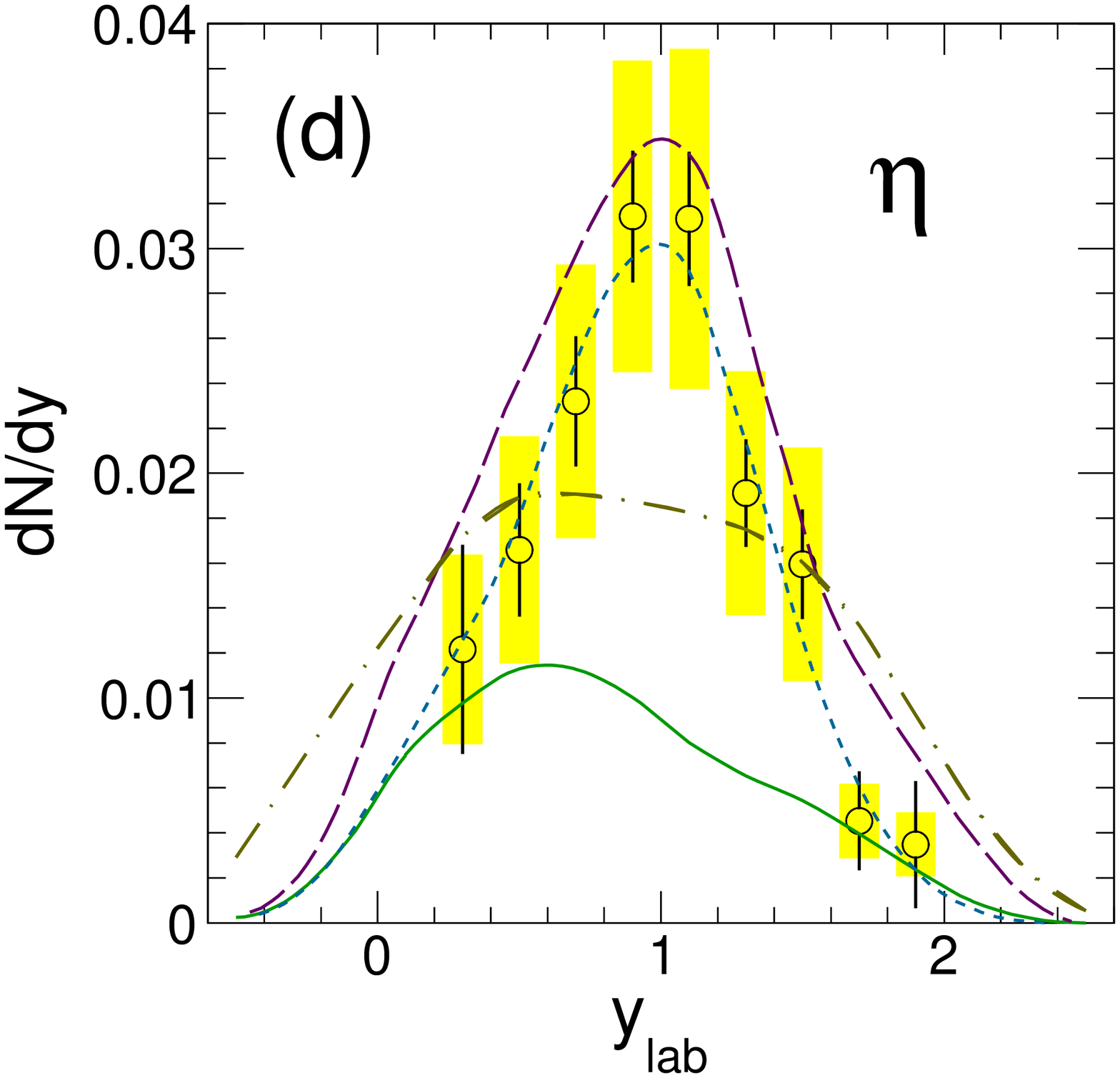}}
  \vspace*{-0.2cm}
  \caption[]{(Color online) (a) $\pi^0$ and (b) $\eta$ transverse momentum distributions
    $dN/dp_{\perp}$ per minimum-bias event in 3.5~\gev\ p+Nb reactions (symbols) compared to
    results of the UrQMD, HSD, and GiBUU transport models.
    For the latter one, also the extended-resonance implementation is shown (see text). 
    All error bars are as in Fig.~\ref{mult_pt}.
    (c) $\pi^0$ (with $p_{\perp} > 0.35$~\gevc) and (d) $\eta$
    rapidity distributions $dN/dy$ per minimum-bias event compared to transport calculations.
    The meaning of the lines is as in (a) and (b),  error bars are as in Fig.~\ref{mult_y}.
  }

  \vspace*{0.2cm}
  \label{transport_pty}
\end{figure*}

Our observation that the rapidity distributions measured in the asymmetric p+Nb system
are centered at values $y_{max} < y_{NN}$ strongly suggests that, beyond first-chance
nucleon-nucleon collisions, secondary reactions, i.e. processes involving multiple
successive interactions of baryons and/or mesons contribute sizably to particle production.
A similar behavior had already been noticed for kaon production in a previous study of
p+Au collisions at comparable bombarding energies \cite{kaos}.  While the observed
phase space population of the eta agrees quite well with a fireball description, this is
questionable for the pion.  Complete thermalization is apparently not reached in the p+Nb
reaction and a transport-theoretical approach is required to model the complex interplay
between reaction dynamics and particle production.  Transport models typically handle
meson and baryon production at energies up to a few \gev\ by resonance excitation and at higher
energies through string fragmentation.  In that respect, our beam energy is particularly
challenging because it is situated in the transition region between these two regimes.

In the following we compare our results with three transport calculations done with either
UrQMD, the Ultrarelativistic Quantum Molecular Dynamics model (version v3.3p1, see \cite{urqmd2}),
GiBUU, the Giessen Boltzmann-Uehling-Uhlenbeck model (version 1.5, see \cite{gibuu2}),
or HSD, the Hadron String Dynamics model (version 2.7, see \cite{hsd2}).
At 3.5~\gev\ bombarding energy, corresponding to $\sqrt{s_{NN}}$ = 3.18~\gev, UrQMD runs
in the resonance regime only whereas HSD switches over to string fragmentation mode at
$\sqrt{s_{NN}}$ = 2.6~\gev.  For GiBUU, on the other hand, we present calculations done
with two different realizations of this model: (i) the original implementation with a smooth
transition to string fragmentation at $\sqrt{s_{NN}}$ = 2.6~\gev\ and (ii) a version
(denoted hereafter by ``ext. res.'') where the resonance region has been extended
up to about $\sqrt{s_{NN}}$ = 3.5~\gev\ \cite{gibuu2}.

\begin{table}[!hbt]

\caption[]{Transport-model calculations of minimum-bias inclusive meson multiplicities
           per p+Nb collision, $N_{\pi^0}$ and $N_{\eta}$, within the accepted rapidity
           range ($0.2<y_{lab}<1.8$) as well as in the full solid angle ($4\pi$).
}

\vspace*{0.2cm}
\begin{center}
\begin{tabular}{l c c c c} 
  \hline
  \hline
\rule{0pt}{2ex} &  \multicolumn{2}{c}{~~~~~$N_{\pi^0}$} & \multicolumn{2}{c}{~~~~~~~$N_{\eta}$} \\
\rule{0pt}{3ex}  Model & $0.2<y<1.8$ & $4\pi$ ~~&~~ $0.2<y<1.8$ & $4\pi$ \\
  \hline
UrQMD v3.3p1  & 0.38 & 0.66 ~~&~~ 0.013 & 0.016 \\
HSD v2.7      & 0.38 & 0.69 ~~&~~ 0.028 & 0.038 \\
GiBUU v1.5    & 0.39 & 0.64 ~~&~~ 0.039 & 0.046 \\
GiBUU ext res & 0.32 & 0.49 ~~&~~ 0.031 & 0.034 \\
  \hline
  \hline

\end{tabular}
\end{center}
\label{modeltable}
\end{table}

Figure~\ref{transport_pty}, panels (a) and (b), shows that $\pi^0$ production is fairly well
described by all models, and this in both observables $p_{\perp}$ and $y$, within the HADES
rapidity and transverse momentum acceptance.  The pion yields reconstructed in the rapidity
range $y_{lab}$ = 0.2 -- 1.8 are in fact reproduced to within 10-25\%
(cf. Table~\ref{multtable} vs. Table~\ref{modeltable}).  The rapidity distributions of
all models are very similar with a slight tendency of being shifted towards target rapidity.
Differences between the various calculations are mostly visible for $p_{\perp}<0.3$~\gevc\
and for $p_{\perp}>1$~\gevc.  Notice also that at low $p_{\perp}$ the standard implementation
of GiBUU behaves more in line with UrQMD than the one with an extended resonance region.
This is somewhat surprising as one would rather expect that this modification of GiBUU
increases the similarity of the two models.

Comparing next in Fig.~\ref{transport_pty}, panels (c) and (d), the calculated and
measured $\eta$ distributions, larger discrepancies between the models do appear.
While UrQMD reproduces quite well the $dN/dp_{\perp}$ shape, it underestimates
the accepted yield by a factor of 2--3 and also misses the $dN/dy$ shape.  Both versions
of GiBUU, on the other hand, do fairly well in describing the $\eta$ rapidity
distribution and integrated yield (see Table~\ref{modeltable}), and its extended-resonance
implementation also possesses the correct transverse-momentum behavior.
The HSD pion and eta yields, finally, do agree fairly well with the data, albeit their
$y$ and $p_{\perp}$ distributions deviate substantially.  The complete lack of data in
the energy range discussed here probably explains why the models tend to perform
worse for $\eta$ production than they do for pions.  We are confident that with the help
of our new results a more detailed theoretical investigation of the relevant
production processes will now be possible.


\section{Summary}
\label{summary}

To summarize, we have presented data on inclusive pion and eta production in the
reaction p+Nb at 3.5~\gev\ kinetic beam energy.   In this study we have used the
photon-conversion method to detect and reconstruct neutral mesons from 4-lepton
final states.  We have demonstrated that with HADES quantitative results on
differential $\pi^0$ and $\eta$ yields can be obtained over a large range of
transverse momentum and rapidity.  Our data provide valuable new input for the
theoretical description of proton-nucleus and nucleus-nucleus collisions in the
few-\gev\ energy regime with respect to both meson dynamics and dilepton emission.
This is exemplified in our comparison with a selection of available transport models
revealing an overall fair to good agreement in various observables.
Together with our previous studies of p+p reactions \cite{hades_p35p, hades_p22p},
the present results provide the required baseline for measurements with
heavy-ion beams at the future FAIR facility.  Indeed, as its central component
--- the SIS100 accelerator --- is designed to provide intense beams of even the
heaviest ions up to 8~\gevu, we will be in the position to isolate unambiguously
those effects induced by the hot and dense baryonic medium.


\acknowledgments {
We thank E.~Bratkovskaya and J.~Weil for providing us with
their latest HSD, respectively GiBUU, transport calculations.
The HADES Collaboration gratefully acknowledges the support
by BMBF grants 06DR9059D, 05P12CRGHE, 06FY171, 06MT238 T5,
and 06MT9156 TP5, by HGF VH-NG-330, by DFG EClust
153, by GSI TMKRUE, by the Hessian LOEWE initiative
through HIC for FAIR (Germany), by EMMI GSI,
by grant GA CR 13-067595 (Czech Rep.),
by grant NN202198639 (Poland),
by grant UCY-10.3.11.12 (Cyprus), by CNRS/IN2P3 (France),
by INFN (Italy), and by EU contracts RII3-CT-2005-515876
and HP2 227431.
}



\end{document}